\documentclass[10pt,twocolumn,letterpaper]{article}

\usepackage[pagenumbers]{cvpr} 

\usepackage{graphicx}
\usepackage{amsmath}
\usepackage{amssymb}
\usepackage{booktabs}
\usepackage{multirow, makecell, arydshln}

\usepackage[pagebackref,breaklinks,colorlinks]{hyperref}

\usepackage[capitalize]{cleveref}
\crefname{section}{Sec.}{Secs.}
\Crefname{section}{Section}{Sections}
\Crefname{table}{Table}{Tables}
\crefname{table}{Tab.}{Tabs.}

\def\cvprPaperID{103} 
\def\confName{CVPR}
\def\confYear{2023}

\begin{document}

\title{Activating More Pixels in Image Super-Resolution Transformer}

\author{
 Xiangyu Chen\textsuperscript{\rm 1,2,3} \space \space 
 Xintao Wang\textsuperscript{\rm 4} \space \space 
 Jiantao Zhou\textsuperscript{\rm 1} \space \space 
 Yu Qiao\textsuperscript{\rm 2,3}   \space \space 
 Chao Dong\textsuperscript{\rm 2,3\dag} \\ 
 \textsuperscript{\rm 1}State Key Laboratory of Internet of Things for Smart City, University of Macau\\
 \textsuperscript{\rm 2}Shenzhen Key Lab of Computer Vision and Pattern Recognition, \\Shenzhen Institute of Advanced Technology, Chinese Academy of Sciences\\
 \textsuperscript{\rm 3}Shanghai Artificial Intelligence Laboratory
 \textsuperscript{\rm 4}ARC Lab, Tencent PCG\\ 
 \url{https://github.com/XPixelGroup/HAT}
}

\maketitle

\renewcommand{\thefootnote}{}
\footnotetext[1]{$^\dag$ Corresponding author.}
\renewcommand{\thefootnote}{}

\begin{abstract}
Transformer-based methods have shown impressive performance in low-level vision tasks, such as image super-resolution. However, we find that these networks can only utilize a limited spatial range of input information through attribution analysis. This implies that the potential of Transformer is still not fully exploited in existing networks. In order to activate more input pixels for better reconstruction, we propose a novel Hybrid Attention Transformer (HAT). 
It combines both channel attention and window-based self-attention schemes, thus making use of their complementary advantages of being able to utilize global statistics and strong local fitting capability.
Moreover, to better aggregate the cross-window information, we introduce an overlapping cross-attention module to enhance the interaction between neighboring window features. In the training stage, we additionally adopt a same-task pre-training strategy to exploit the potential of the model for further improvement. Extensive experiments show the effectiveness of the proposed modules, and we further scale up the model to demonstrate that the performance of this task can be greatly improved. Our overall method significantly outperforms the state-of-the-art methods by more than \textbf{1dB}. 
\end{abstract}

\vspace{-10pt}
\section{Introduction}
Single image super-resolution (SR) is a classic problem in computer vision and image processing. It aims to reconstruct a high-resolution image from a given low-resolution input. Since deep learning has been successfully applied to the SR task~\cite{srcnn_eccv}, numerous methods based on the convolutional neural network (CNN) have been proposed~\cite{srcnn_tpami,fsrcnn,edsr,rcan,rdn,san,classsr,bsrn} and almost dominate this field in the past few years. 
Recently, due to the success in natural language processing, Transformer~\cite{transformer} has attracted the attention of the computer vision community. After making rapid progress on high-level vision tasks~\cite{vit,swin_t,pvt}, Transformer-based methods are also developed for low-level vision tasks~\cite{ipt,uformer,restormer}, as well as for SR~\cite{edt,swinir}. Especially, a newly designed network, SwinIR~\cite{swinir}, obtains a breakthrough improvement in this task. 

\begin{figure}[!t]
\vspace{-0.1cm}
\centering
\includegraphics[width=1\linewidth]{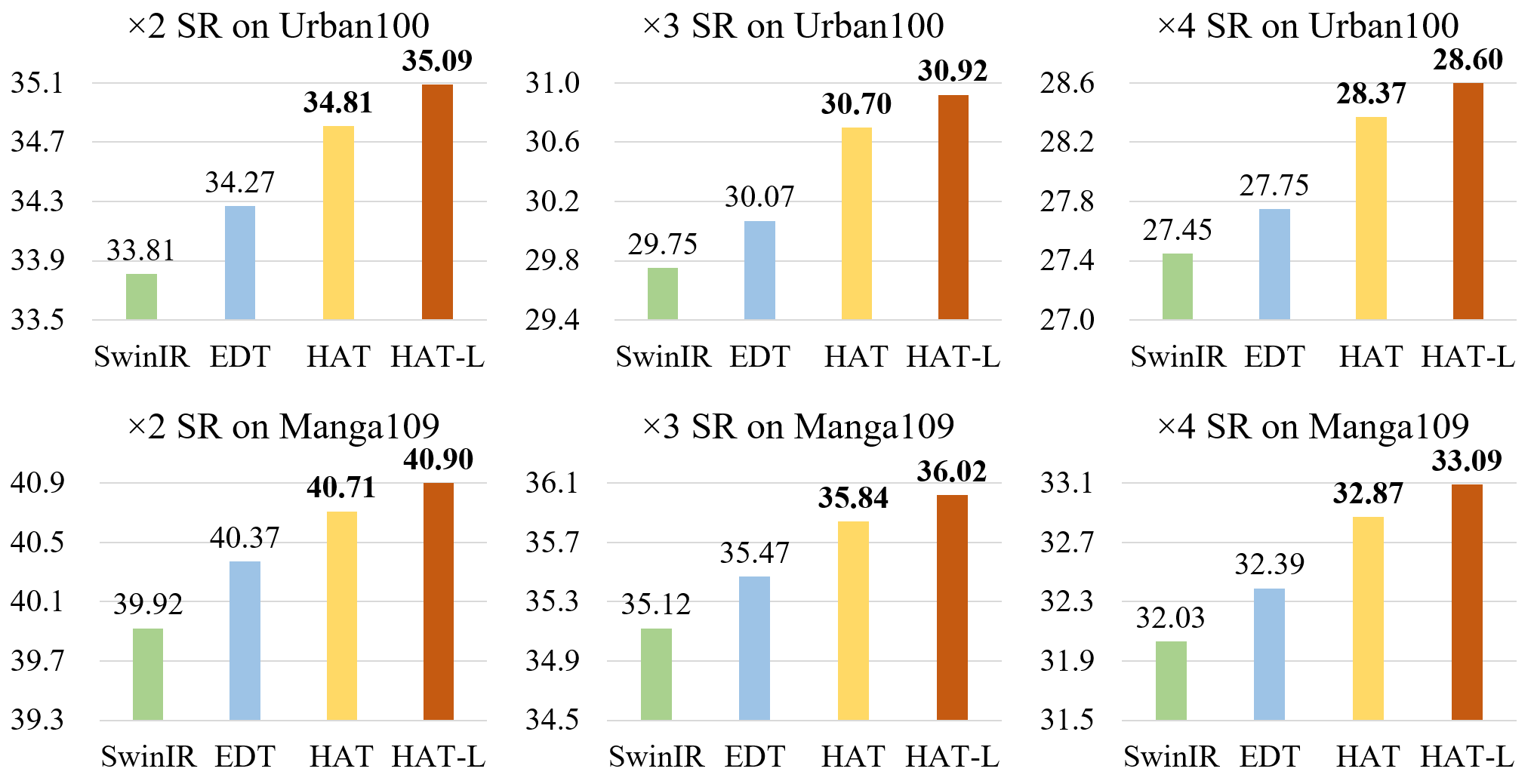}
\vspace{-0.6cm}
\caption{Performance comparison on PSNR(dB) of the proposed HAT with the state-of-the-art methods SwinIR~\cite{swinir} and EDT~\cite{edt}. HAT-L represents a larger variant of HAT. Our approach can surpass the state-of-the-art methods by 0.3dB$\sim$1.2dB.}
\vspace{-0.6cm}
\label{performance}
\end{figure}

Despite the success, ``why Transformer is better than CNN'' remains a mystery. An intuitive explanation is that this kind of network can benefit from the self-attention mechanism and utilize long-range information. 
Thus, we employ the attribution analysis method LAM~\cite{lam} to examine the involved range of utilized information for reconstruction in SwinIR. Interestingly, we find that SwinIR does NOT exploit more input pixels than CNN-based methods (\textit{e.g.}, RCAN~\cite{rcan}) in super-resolution, as shown in Fig.~\ref{lam}.
%
Besides, although SwinIR obtains higher quantitative performance on average, it produces inferior results to RCAN in some samples, due to the limited range of utilized information. 
These phenomena illustrate that Transformer has a stronger ability to model local information, but the range of its utilized information needs to be expanded.
In addition, we also find that blocking artifacts would appear in the intermediate features of SwinIR, as depicted in Fig.~\ref{feature}. It demonstrates that the shift window mechanism cannot perfectly realize cross-window information interaction. 

To address the above-mentioned limitations and further develop the potential of Transformer for SR, we propose a Hybrid Attention Transformer, namely HAT. Our HAT combines channel attention and self-attention schemes, in order to take advantage of the former's capability in using global information and the powerful representative ability of the latter. 
Besides, we introduce an overlapping cross-attention module to achieve more direct interaction of adjacent window features. Benefiting from these designs, our model can activate more pixels for reconstruction and thus obtains significant performance improvement.  

Since Transformers do not have an inductive bias like CNNs, large-scale data pre-training is important to unlock the potential of such models. 
In this work, we provide an effective \textit{same-task pre-training} strategy. Different from IPT~\cite{ipt} using multiple restoration tasks for pre-training and EDT~\cite{edt} using multiple degradation levels for pre-training, we directly perform pre-training using large-scale dataset on the same task. We believe that large-scale data is what really matters for pre-training, and experimental results also show the superiority of our strategy. Equipped with the above designs, HAT can surpass the state-of-the-art methods by a huge margin (0.3dB$\sim$1.2dB), as shown in Fig.~\ref{performance}. 






\textbf{Contributions: 1)} We design a novel Hybrid Attention Transformer (HAT) that combines self-attention, channel attention and a new overlapping cross-attention to activate more pixels for better reconstruction. \textbf{2)} We propose an effective same-task pre-training strategy to further exploit the potential of SR Transformer and show the importance of large-scale data pre-training for the task. \textbf{3)} Our method achieves state-of-the-art performance. By further scaling up HAT to build a big model, we greatly extend the performance upper bound of the SR task. 



\section{Related Work}

\subsection{Deep Networks for Image SR} 

Since SRCNN~\cite{srcnn_eccv} first introduces deep convolution neural networks (CNNs) to the image SR task and obtains superior performance over conventional SR methods, numerous deep networks~\cite{srcnn_tpami,fsrcnn,pixelshuffle,vdsr,edsr,rdn,rcan,san,han,nlsn,swinir,edt} have been proposed for SR to further improve the reconstruction quality. For instance, many methods apply more elaborate convolution module designs, such as residual block~\cite{srgan,edsr} and dense block~\cite{esrgan,rdn}, to enhance the model representation ability. Several works explore more different frameworks like recursive neural network~\cite{drcn,drrn} and graph neural network~\cite{ignn}. To improve perceptual quality, \cite{srgan,esrgan,ranksrgan,realesrgan} introduce adversarial learning to generate more realistic results. By using attention mechanism, \cite{rcan,san,rnan,nlrn,han,nlsn} achieve further improvement in terms of reconstruction fidelity. Recently, a series of Transformer-based networks~\cite{ipt,swinir,edt} are proposed and constantly refresh the state-of-the-art of SR task, showing the powerful representation ability of Transformer.  

To better understand the working mechanisms of SR networks, several works are proposed to analyze and interpret the SR networks. LAM~\cite{lam} adopts the integral gradient method to explore which input pixels contribute most to the final performance. DDR~\cite{ddr} reveals the deep semantic representations in SR networks based on deep feature dimensionality reduction and visualization. FAIG~\cite{faig} aims to find discriminative filters for specific degradations in blind SR. RDSR\cite{dropoutSR} introduces channel saliency map to demonstrate that Dropout can help prevent co-adapting for real-SR networks. SRGA~\cite{srga} aims to evaluate the generalization ability of SR methods. In this work, we exploit LAM~\cite{lam} to analyse and understand the behavior of SR networks.

\subsection{Vision Transformer}
Recently, Transformer~\cite{transformer} has attracted the attention of computer vision community due to its success in the field of natural language processing. A series of Transformer-based methods~\cite{vit,localvit,cvt,pvt,swin_t,shufflet,cswin,twins,palet,ceit,uniformer,moa} have been developed for high-level vision tasks, including image classification~\cite{swin_t,vit,localvit,ramachandran2019studying,vaswani2021scaling}, object detection~\cite{liu2020deep,touvron2021training,swin_t,detr,twins}, segmentation~\cite{wu2020visual,pvt,dat,cao2021swin}, \textit{etc}. Although vision Transformer has shown its superiority on modeling long-range dependency~\cite{vit,cka}, there are still many works demonstrating that the convolution can help Transformer achieve better visual representation~\cite{cvt,vitc,hrformer,ceit,uniformer}. Due to the impressive performance, Transformer has also been introduced for low-level vision tasks~\cite{ipt,uformer,swinir,vsrt,maxim,restormer,vrt,edt}. Specifically, IPT~\cite{ipt} develops a ViT-style network and introduces multi-task pre-training for image processing. SwinIR~\cite{swinir} proposes an image restoration Transformer based on \cite{swin_t}. VRT\cite{vrt} introduces Transformer-based networks to video restoration. EDT\cite{edt} adopts self-attention mechanism and multi-related-task pre-training strategy to further refresh the state-of-the-art of SR. However, existing works still cannot fully exploit the potential of Transformer, while our method can activate more input pixels for better reconstruction.

\section{Methodology}
\subsection{Motivation}
\label{Motivation}

\begin{figure}[!t]
\centering
\includegraphics[width=0.98\linewidth]{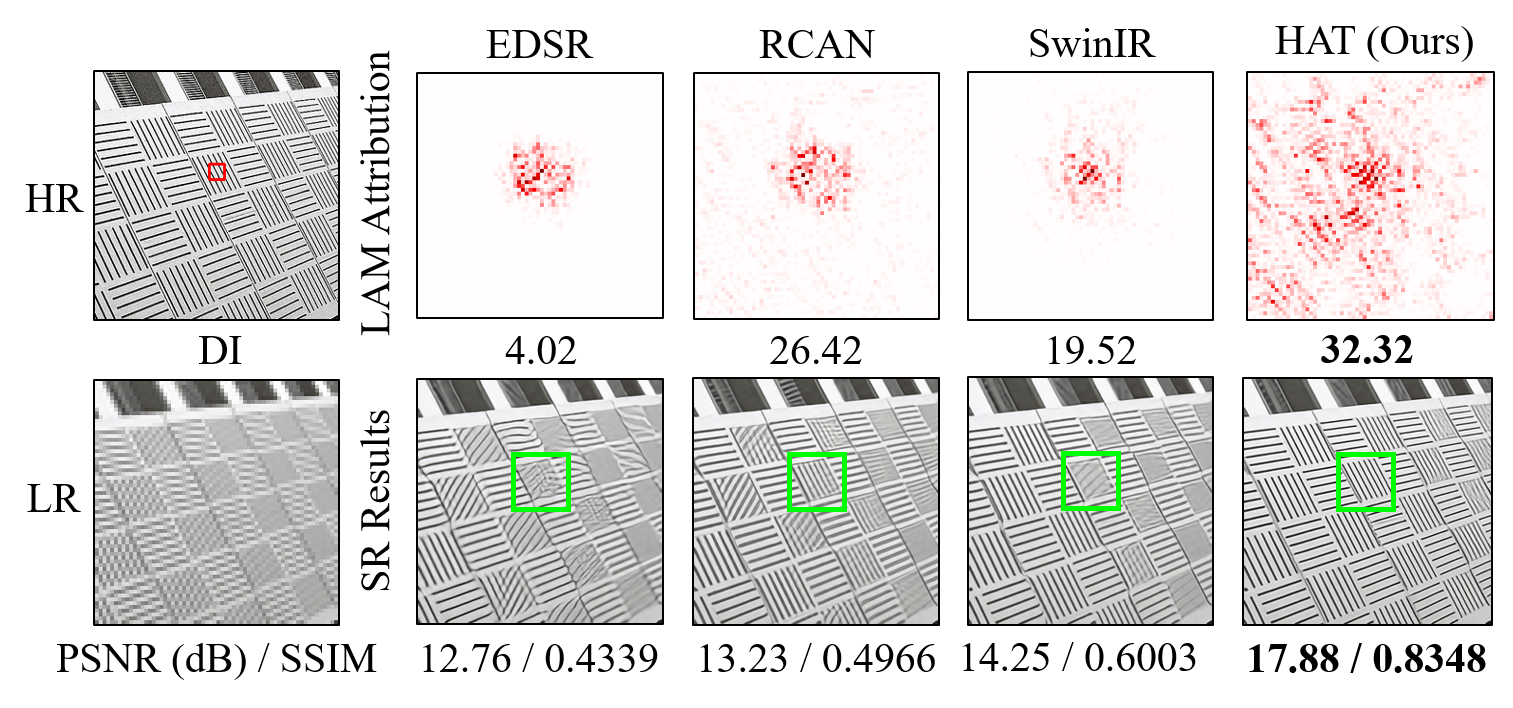}
\vspace{-0.3cm}
\caption{LAM~\cite{lam} results for different networks. The LAM attribution reflects the importance of each pixel in the input LR image when reconstructing the patch marked with a box.
Diffusion index (DI)~\cite{lam} reflects the range of involved pixels. A higher DI represents a wider range of utilized pixels. The results indicate that SwinIR utilizes less information compared to RCAN, while HAT uses the most pixels for reconstruction.}
\label{lam}
\vspace{-0.4cm}
\end{figure}

Swin Transformer~\cite{swin_t} has already presented excellent performance in image super-resolution~\cite{swinir}. Then we are eager to know what makes it work better than CNN-based methods. To reveal its working mechanisms, we resort to a diagnostic tool -- LAM~\cite{lam}, which is an attribution method designed for SR. With LAM, we could tell which input pixels contribute most to the selected region. As shown in Fig.~\ref{lam}, the red marked points are informative pixels that contribute to the reconstruction. Intuitively, the more information is utilized, the better performance can be obtained. This is true for CNN-based methods, as comparing RCAN~\cite{rcan} and EDSR~\cite{edsr}. However, for the Transformer-based method -- SwinIR, its LAM does not show a larger range than RCAN. This is in contradiction with our common sense, but could also provide us with additional insights. First, it implies that SwinIR has a much stronger mapping ability than CNN, and thus could use less information to achieve better performance. Second, SwinIR may restore wrong textures due to the limited range of utilized pixels, and we think it can be further improved if it could exploit more input pixels. Therefore, we aim to design a network that can take advantage of similar self-attention while activating more pixels for reconstruction. As depicted in Fig.~\ref{lam}, our HAT can see pixels almost all over the image and restore correct and clear textures. 


\begin{figure}[!t]
\centering
\includegraphics[width=0.98\linewidth]{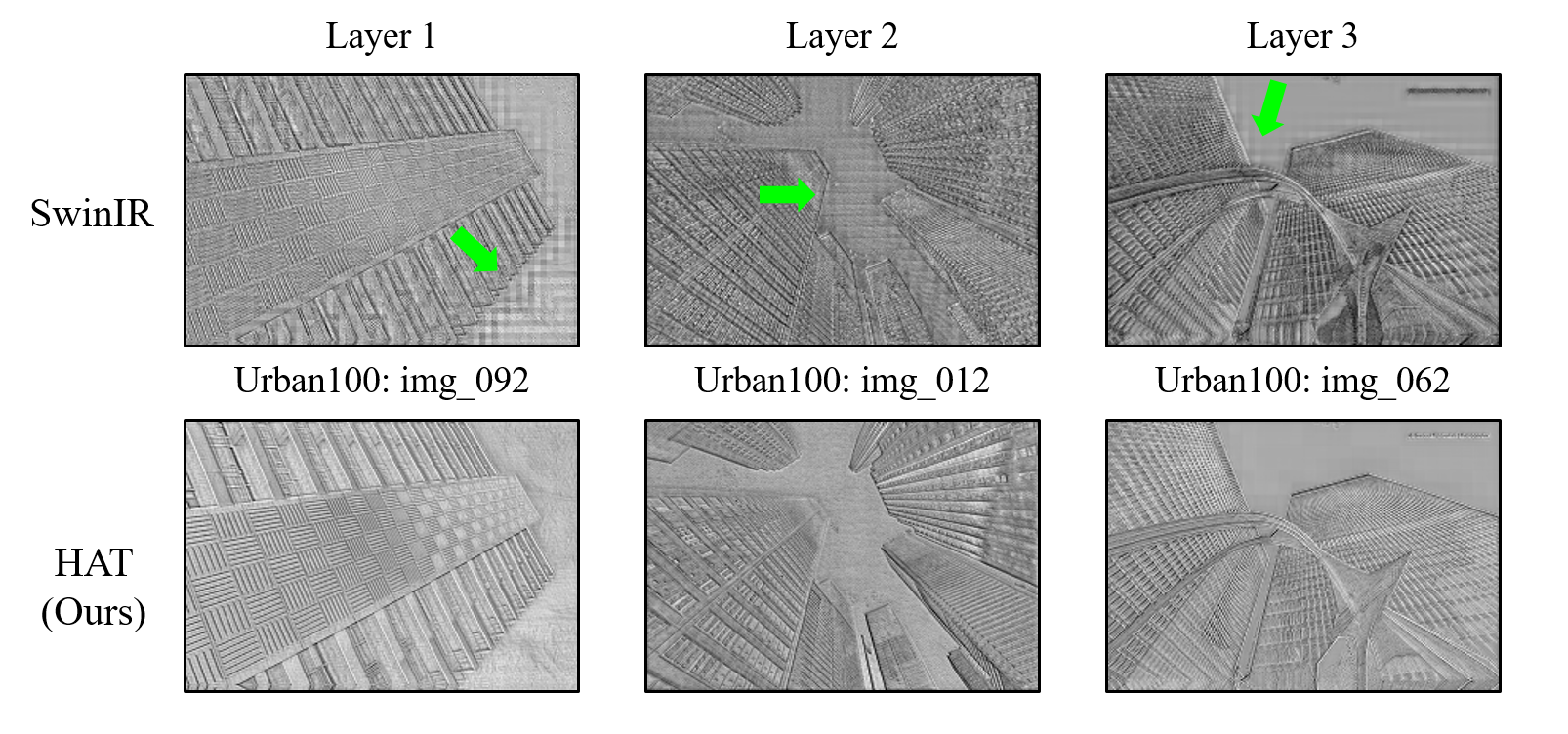}
\vspace{-0.4cm}
\caption{The blocking artifacts appear in the intermediate features of SwinIR~\cite{swinir}. “Layer N ” represents the intermediate features after the $N_{th}$ layer (\textit{i.e.}, RSTB in SwinIR and RHAG in HAT.)}
\label{feature}
\vspace{-0.6cm}
\end{figure}

Besides, we can observe obvious blocking artifacts in the intermediate features of SwinIR, as shown in Fig.~\ref{feature}. These artifacts are caused by the window partition mechanism, which suggests that the shifted window mechanism is inefficient to build the cross-window connection.
Some works for high-level vision tasks~\cite{cswin,shufflet,palet,moa} also point out that enhancing the connection among windows can improve the window-based self-attention methods.
Thus, we strengthen cross-window information interactions when designing our approach and the blocking artifacts in the intermediate features obtained by HAT are significantly alleviated. 


\subsection{Network Architecture}

\subsubsection{The Overall Structure}

\begin{figure*}[!t]
\centering
\includegraphics[width=0.95\linewidth]{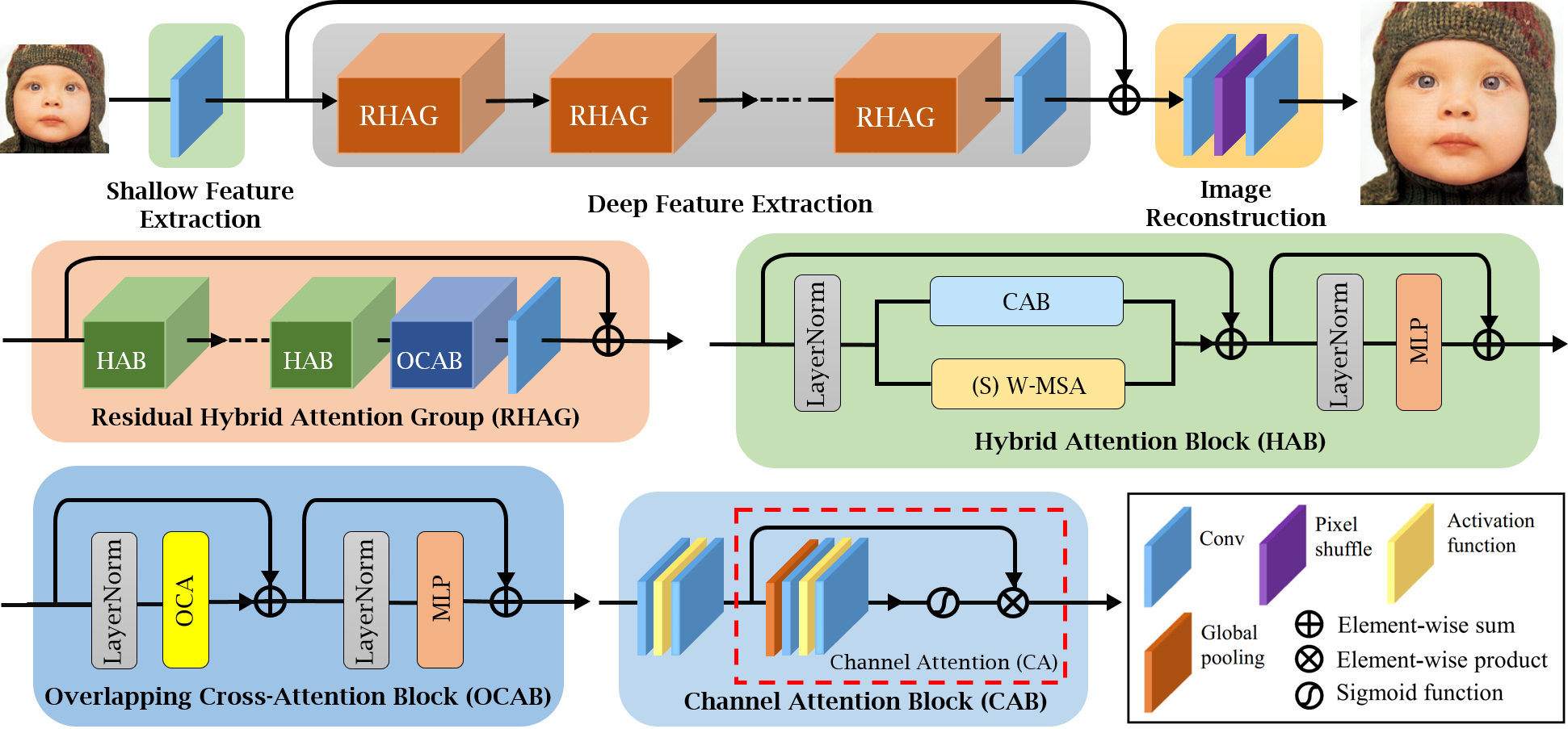}
\vspace{-0.1cm}
\caption{The overall architecture of HAT and the structure of RHAG and HAB.}
\label{Network Structure}
\vspace{-0.5cm}
\end{figure*}

As shown in Fig.~\ref{Network Structure}, the overall network consists of three parts, including shallow feature extraction, deep feature extraction and image reconstruction. The architecture design is widely used in previous works \cite{rcan,swinir}. Specifically, for a given low-resolution (LR) input $I_{LR}\in\mathbb{R}^{H\times W\times C_{in}}$, we first exploit one convolution layer to extract the shallow feature $F_0\in\mathbb{R}^{H\times W\times C}$, where $C_{in}$ and $C$ denote the channel number of the input and the intermediate feature. Then, a series of residual hybrid attention groups (RHAG) and one $3\times 3$ convolution layer $H_{Conv}(\cdot)$ are utilized to perform the deep feature extraction. After that, we add a global residual connection to fuse shallow features $F_0$ and deep features $F_D\in\mathbb{R}^{H\times W\times C}$, and then reconstruct the high-resolution result via a reconstruction module. As depicted in Fig.~\ref{Network Structure}, each RHAG contains several hybrid attention blocks (HAB), an overlapping cross-attention block (OCAB) and a $3\times 3$ convolution layer with a residual connection. For the reconstruction module, the pixel-shuffle method~\cite{pixelshuffle} is adopted to up-sample the fused feature. We simply use $L_1$ loss to optimize the network parameters.

\subsubsection{Hybrid Attention Block (HAB)} 
\label{HAB}
As shown in Fig.~\ref{lam}, more pixels are activated when channel attention is adopted, as global information is involved to calculate the channel attention weights. Besides, many works illustrate that convolution can help Transformer get better visual representation or achieve easier optimization~\cite{cvt,vitc,ceit,uniformer,spach}. Therefore, we incorporate a channel attention-based convolution block into the standard Transformer block to enhance the representation ability of the network. As demonstrated in Fig.~\ref{Network Structure}, a channel attention block (CAB) is inserted into the standard Swin Transformer block after the first LayerNorm (LN) layer in parallel with the window-based multi-head self-attention (W-MSA) module. Note that shifted window-based self-attention (SW-MSA) is adopted at intervals in consecutive HABs similar to \cite{swin_t,swinir}. 
To avoid the possible conflict of CAB and MSA on optimization and visual representation, a small constant $\alpha$ is multiplied to the output of CAB. For a given input feature $X$, the whole process of HAB is computed as 
\begin{gather}
  X_N={\rm LN}(X), \notag \\
  X_M={\rm \text{(S)W-MSA}}(X_N)+\alpha {\rm CAB} (X_N)+X, \\
  Y={\rm MLP}({\rm LN}(X_M))+X_M, \notag 
\end{gather}
where $X_N$ and $X_M$ denote the intermediate features. $Y$ represents the output of HAB. 
Especially, we treat each pixel as a token for embedding (\textit{i.e.}, set patch size as 1 for patch embedding following~\cite{swinir}). MLP denotes a multi-layer perceptron. For  calculation of the self-attention module, given an input feature of size $H\times W\times C$, it is first partitioned into $\frac{HW}{M^2}$ local windows of size $M\times M$, then self-attention is calculated inside each window. For a local window feature $X_W\in\mathbb{R}^{M^2\times C}$, the \textit{query}, \textit{key} and \textit{value} matrices are computed by linear mappings as $Q$, $K$ and $V$. Then the window-based self-attention is formulated as 
\begin{equation}
    {\rm Attention}(Q,K,V)={\rm SoftMax}(QK^T/\sqrt{d}+B)V, \label{attn}
\end{equation}
where $d$ represents the dimension of \textit{query}/\textit{key}. $B$ denotes the relative position encoding and is calculated as \cite{transformer}. Note that we use a large window size to compute self-attention, since we find it significantly enlarges the range of used pixels, as depicted in Sec.\ref{cmp_win_size}. Besides, to build the connections between neighboring non-overlapping windows, we also utilize the shifted window partitioning approach~\cite{swin_t} and set the shift size to half of the window size. 

A CAB consists of two standard convolution layers with a GELU activation~\cite{GELU} and a channel attention (CA) module, as shown in Fig.~\ref{Network Structure}. Since the Transformer-based structure often requires a large number of channels for token embedding, directly using convolutions with constant width incurs a large computation cost. Thus, we compress the channel numbers of the two convolution layers by a constant $\beta$. For an input feature with $C$ channels, the channel number of the output feature after the first convolution layer is squeezed to $\frac{C}{\beta}$, then the feature is expanded to $C$ channels through the second layer. Next, a standard CA module ~\cite{rcan} is exploited to adaptively rescale channel-wise features. 


\begin{figure}[!t]
\centering
\includegraphics[width=0.95\linewidth]{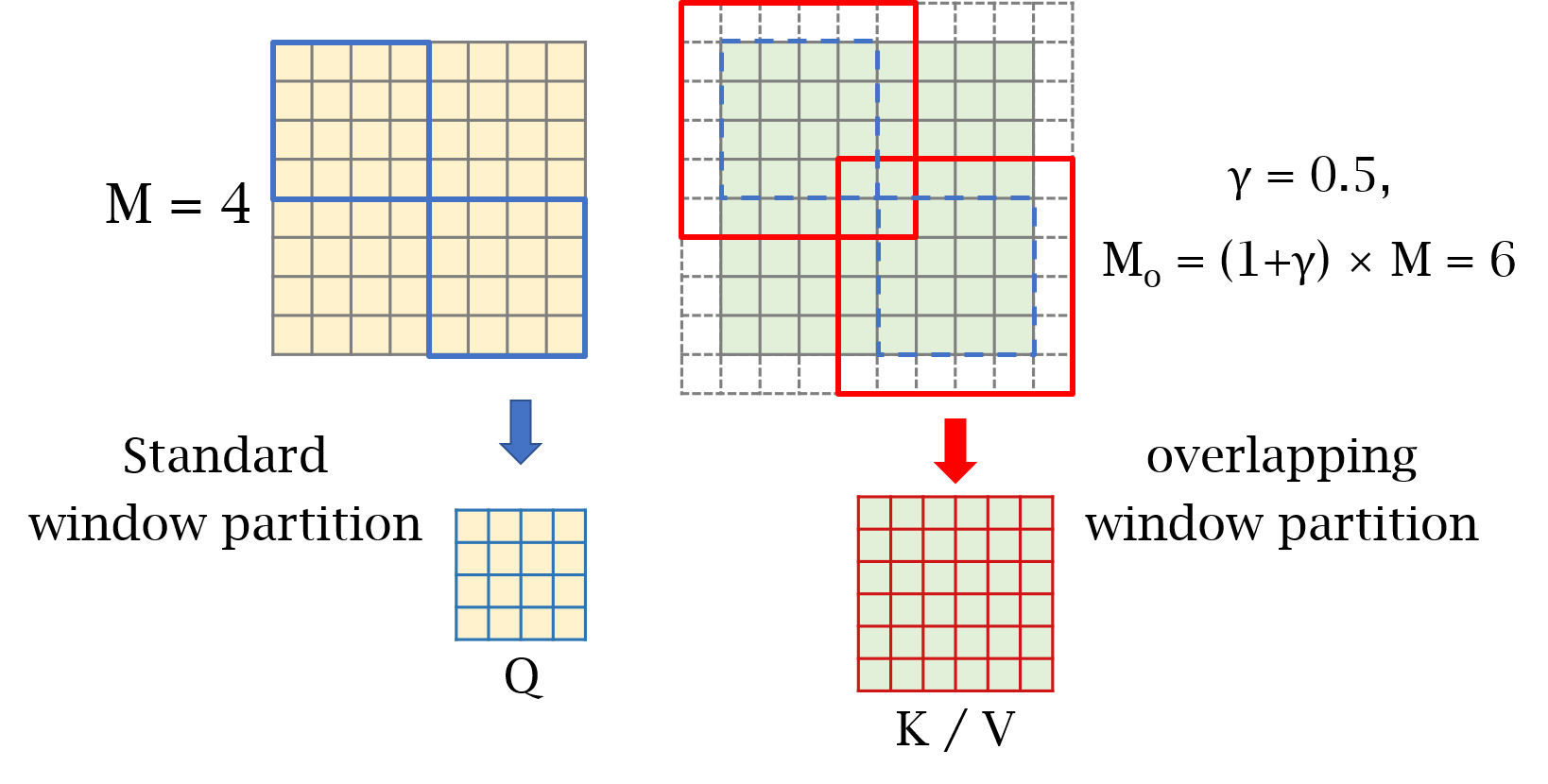}
\vspace{-0.3cm}
\caption{The overlapping window partition for OCA.}
\vspace{-0.5cm}
\label{OCA}
\end{figure}

\subsubsection{Overlapping Cross-Attention Block (OCAB)}
\label{ocab}
We introduce OCAB to directly establish cross-window connections and enhance the representative ability for the window self-attention. Our OCAB consists of an overlapping cross-attention (OCA) layer and an MLP layer similar to the standard Swin Transformer block~\cite{swin_t}.  But for OCA, as depicted in Fig.~\ref{OCA}, we use different window sizes to partition the projected features. Specifically, for the $X_Q,X_K,X_V\in\mathbb{R}^{H\times W\times C}$ of the input feature $X$, $X_Q$ is partitioned into $\frac{HW}{M^2}$ non-overlapping windows of size ${M}\times {M}$, while $X_K,X_V$ are unfolded to $\frac{HW}{M^2}$ overlapping windows of size ${M_o}\times {M_o}$. It is calculated as
\begin{equation}
    M_o=(1+\gamma)\times M,
\end{equation}
where $\gamma$ is a constant to control the overlapping size. To better understand this operation, the standard window partition can be considered as a sliding partition with the kernel size and the stride both equal to the window size $M$. In contrast, the overlapping window partition can be viewed as a sliding partition with the kernel size equal to $ M_o$, while the stride is equal to $M$. Zero-padding with size $\frac{\gamma M}{2}$ is used to ensure the size consistency of overlapping windows. The attention matrix is calculated as Equ.~\ref{attn}, and the relative position bias $B\in\mathbb{R}^{M\times M_o}$ is also adopted. Unlike WSA whose \textit{query}, \textit{key} and \textit{value} are calculated from the same window feature, OCA computes \textit{key}/\textit{value} from a larger field where more useful information can be utilized for the \textit{query}. 
Note that although Multi-resolution Overlapped Attention (MOA) module in~\cite{moa} performs similar overlapping window partition, our OCA is fundamentally different from MOA, since MOA calculates global attention using window features as tokens while OCA computes cross-attention inside each window feature using pixel token. 

\subsection{The Same-task Pre-training}
Pre-training is proven effective on many high-level vision tasks~\cite{vit,bao2021beit,he2022masked}. Recent works~\cite{ipt,edt} also demonstrate that pre-training is beneficial to low-level vision tasks. IPT~\cite{ipt} emphasizes the use of various low-level tasks, such as denoising, deraining, super-resolution and \textit{etc.}, while EDT~\cite{edt} utilizes different degradation levels of a specific task to do pre-training. These works focus on investigating the effect of multi-task pre-training for a target task. In contrast, we directly perform pre-training on a larger-scale dataset (\textit{i.e.}, ImageNet~\cite{imagenet}) based on the same task, showing that the effectiveness of pre-training depends more on the scale and diversity of data. For example, when we want to train a model for $\times 4$ SR, we first train a $\times 4$ SR model on ImageNet, then fine-tune it on the specific dataset, such as DF2K. The proposed strategy, namely \textit{same-task pre-training}, is simpler while bringing more performance improvements. It is worth mentioning that sufficient training iterations for pre-training and an appropriate small learning rate for fine-tuning are very important for the effectiveness of the pre-training strategy. We think that it is because Transformer requires more data and iterations to learn general knowledge for the task, but 
needs a small learning rate for fine-tuning to avoid overfitting to the specific dataset.

\section{Experiments}

\subsection{Experimental Setup}
We use DF2K (DIV2K~\cite{div2k}+Flicker2K~\cite{flicker2k}) dataset as the training dataset, since we find that using only DIV2K will lead to overfitting. When utilizing pre-training, we adopt ImageNet~\cite{imagenet} following \cite{ipt,edt}. For the structure of HAT, we keep the depth and width the same as SwinIR. Specifically, the RHAG number and HAB number are both set to 6. The channel number is set to 180. The attention head number and window size are set to 6 and 16 for both (S)W-MSA and OCA. For the hyper-parameters of the proposed modules, we set the weighting factor in HAB ($\alpha$), the squeeze factor between two convolutions in CAB ($\beta$), and the overlapping ratio of OCA ($\gamma$) as 0.01, 3 and 0.5. For the large variant HAT-L, we directly double the depth of HAT by increasing the RHAG number from 6 to 12. We also provide a small version HAT-S with fewer parameters and similar computation to SwinIR. In HAT-S, the channel number is set to 144 and the depth-wise convolution is used in CAB. Five benchmark datasets including Set5~\cite{set5}, Set14~\cite{set14}, BSD100~\cite{bsd100}, Urban100~\cite{urban100} and Manga109~\cite{manga109} are used to evaluate the methods. For the quantitative metrics, PSNR and SSIM (calculated on the Y channel) are reported. More training details can refer to the \textit{supp.} file.

\subsection{Effects of different window sizes}
\label{cmp_win_size}
\begin{figure}[!t]
\centering
\includegraphics[width=0.95\linewidth]{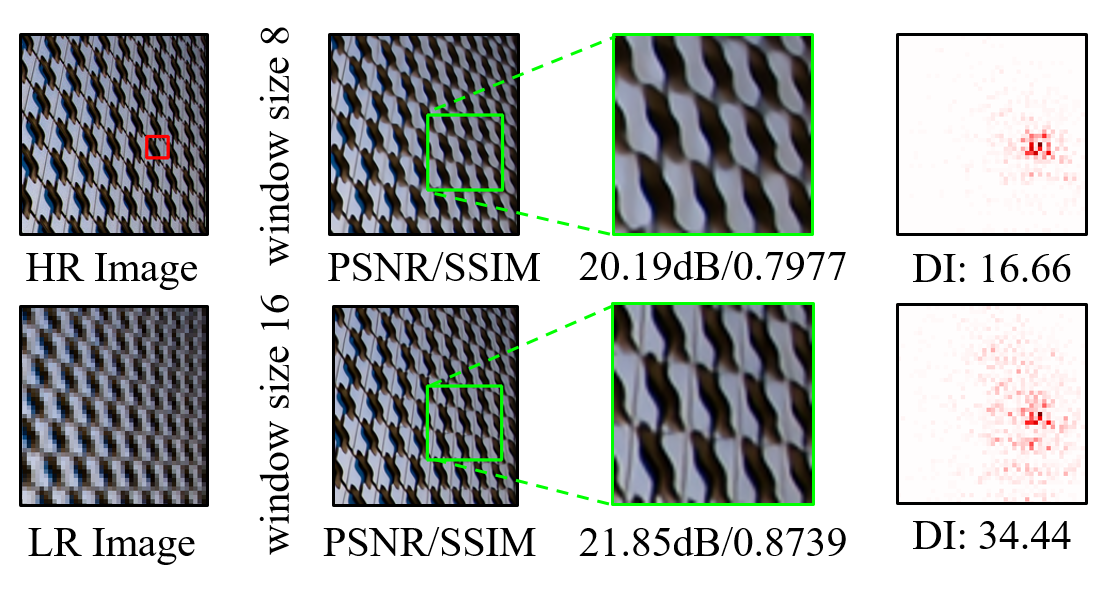}
\vspace{-0.3cm}
\caption{Qualitative comparison of different window sizes.}
\vspace{-0.5cm}
\label{win_size_cmp_fig}
\end{figure}

As discussed in Sec. \ref{Motivation}, activating more input pixels for SR tends to achieve better performance. 
Enlarging window size for the window-based self-attention is an intuitive way to realize the goal.
In \cite{edt}, the authors investigate the effects of different window sizes. However, they conduct experiments based on the shifted cross local attention and only explore the window size up to 12$\times$12. We further explore how the window size of self-attention influences the representation ability. To eliminate the influence of our newly-introduced blocks, we conduct the following experiments directly on SwinIR. As shown in Tab.~\ref{win_size_cmp_lab}, the model with a large window size of 16$\times$16 obtains better performance, especially on the Urban100. We also provide the qualitative comparison in Fig.~\ref{win_size_cmp_fig}. For the red marked patch, the model with window size of 16 utilizes much more input pixels than the model with window size of 8. The quantitative performance of the reconstructed results also demonstrates the effectiveness of large window size. Based on this conclusion, we directly use window size 16 as our default setting. 

\begin{table}[!t]
\center
\begin{center}
\caption{Quantitative comparison on PSNR(dB) of different window sizes.}
\vspace{-0.2cm}
\label{win_size_cmp_lab}
\setlength{\tabcolsep}{1.5mm}{
\begin{tabular}{c|ccccc} 
\hline 
Size & Set5 & Set14 & BSD100 & Urban100 & Manga109\\
\hline 
(8,8) & 32.88 & 29.09 & 27.92 & 27.45 & 32.03\\
(16,16) & 32.97 & 29.12 & 27.95 & 27.81 & 32.15\\
\hline 
\end{tabular}
}
\end{center}
\vspace{-0.2cm}
\end{table}

\subsection{Ablation Study}
\begin{figure}[!t]
\centering
\includegraphics[width=1.0\linewidth]{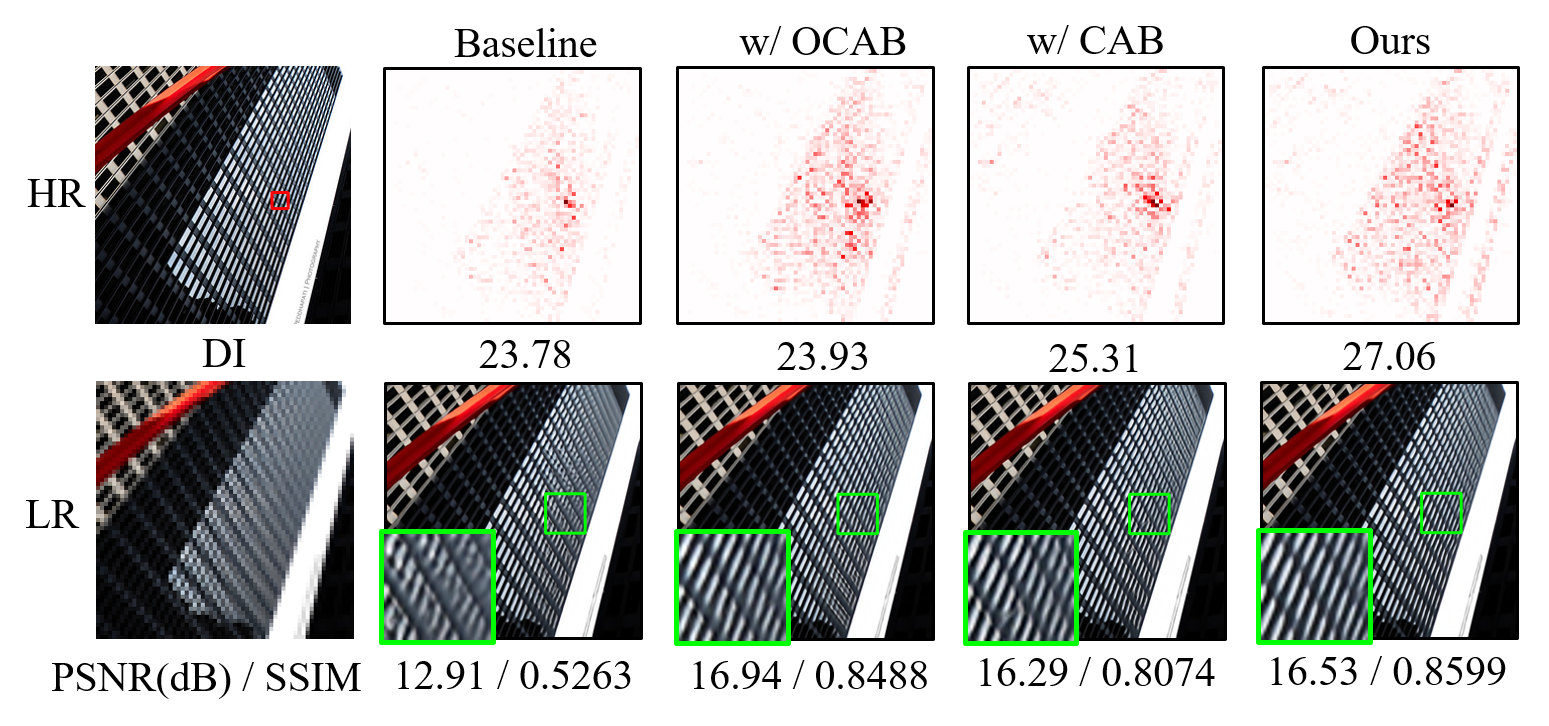}
\vspace{-0.7cm}
\caption{Ablation study on the proposed OCAB and CAB. 
}
\label{ablation_study_fig}
\vspace{-0.7cm}
\end{figure}

\noindent
\textbf{Effectiveness of OCAB and CAB.}
We conduct experiments to demonstrate the effectiveness of the proposed CAB and OCAB. The quantitative performance reported on the Urban100 dataset for $\times4$ SR is shown in Tab.~\ref{ablation_study_tab}. 
Compared with the baseline results, both OCAB and CAB bring the performance gain of 0.1dB. Benefiting from the two modules, the model obtains a further performance improvement of 0.16dB. We also provide qualitative comparison to further illustrate the influence of OCAB and CAB, as presented in Fig.~\ref{ablation_study_fig}. We can observe that the model with OCAB has a larger scope of the utilized pixels and generate better-reconstructed results. When CAB is adopted, the used pixels even expand to almost the full image. Moreover, the result of our method with OCAB and CAB obtains the highest DI\cite{lam}, which means our method utilizes the most input pixels. Although it obtains a little lower performance than w/OCAB, our method gets the highest SSIM and reconstructs the clearest textures.

\begin{table}[!t]
\center
\begin{center}
\caption{Ablation study on the proposed OCAB and CAB.}
\vspace{-0.2cm}
\label{ablation_study_tab}
\setlength{\tabcolsep}{2mm}{
\begin{tabular}{c|cccc} 
\hline 
~ & \multicolumn{4}{c}{Baseline} \\
\hline 
OCAB & \text{\sffamily X} & \checkmark & \text{\sffamily X} & \checkmark\\
CAB & \text{\sffamily X} & \text{\sffamily X} & \checkmark & \checkmark\\
\hline 
PSNR & 27.81dB & 27.91dB & 27.91dB & 27.97dB\\
\hline 
\end{tabular}
}
\end{center}
\vspace{-0.3cm}
\end{table}

\noindent
\textbf{Effects of different designs of CAB.} 
We conduct experiments to explore the effects of different designs of CAB. First, we investigate the influence of channel attention. As shown in Tab. \ref{CAB_structure}, the model using CA achieves a performance gain of 0.05dB compared to the model without CA. It demonstrates the effectiveness of the channel attention in our network. 
We also conduct experiments to explore the effects of the weighting factor $\alpha$ of CAB. As presented in the manuscript Sec. \ref{HAB}, $\alpha$ is used to control the weight of CAB features for feature fusion. A larger $\alpha$ means a larger weight of features extracted by CAB and $\alpha=0$ represents CAB is not used. As shown in Tab.~\ref{weighting_factor}, the model with $\alpha$ of 0.01 obtains the best performance. It indicates that CAB and self-attention may have potential issue in optimization, while a small weighting factor for the CAB branch can suppress this issue for the better combination.

\noindent
\textbf{Effects of the overlapping ratio.}
In OCAB, we set a constant $\gamma$ to control the overlapping size for the overlapping cross-attention. To explore the effects of different overlapping ratios, we set a group of $\gamma$ from 0 to 0.75 to examine the performance change, as shown in Tab. \ref{overlapping size}. Note that $\gamma=0$ means a standard Transformer block. It can be found that the model with $\gamma=0.5$ performs best. In contrast, when $\gamma$ is set to 0.25 or 0.75, the model has no obvious performance gain or even has a performance drop. It illustrates that inappropriate overlapping size cannot benefit the interaction of neighboring windows. 

\begin{table}[!t]
\center
\begin{center}
\caption{Effects of the channel attention (CA) module in CAB.}
\vspace{-0.2cm}
\label{CAB_structure}
\setlength{\tabcolsep}{2.2mm}{
\begin{tabular}{c|cc} 
\hline 
Structure & w/o CA & w/ CA\\
\hline 
PSNR / SSIM & 27.92dB / 0.8362 & 27.97dB / 0.8367\\
\hline 
\end{tabular}
}
\end{center}
\vspace{-0.1cm}
\end{table}

\begin{table}[!t]
\center
\begin{center}
\caption{Effects of the weighting factor $\alpha$ in CAB.}
\vspace{-0.2cm}
\label{weighting_factor}
\setlength{\tabcolsep}{2mm}{
\begin{tabular}{c|cccc} 
\hline 
$\alpha$ & 0 & 1 & 0.1 & 0.01\\
\hline 
PSNR & 27.81dB & 27.86dB & 27.90dB & 27.97dB\\
\hline 
\end{tabular}
}
\end{center}
\vspace{-0.1cm}
\end{table}

\begin{table}[!t]
\center
\begin{center}
\caption{Ablation study on the overlapping ratio of OCAB.}
\vspace{-0.2cm}
\label{overlapping size}
\setlength{\tabcolsep}{2.5mm}{
\begin{tabular}{c|cccc} 
\hline 
$\gamma$ & 0 & 0.25 & 0.5 & 0.75\\
\hline 
PSNR & 27.85dB & 27.81dB & 27.91dB & 27.86dB\\
\hline 
\end{tabular}
}
\end{center}
\vspace{-0.4cm}
\end{table}

\begin{table*}[!ht]
\center
\begin{center}
\caption{Quantitative comparison with state-of-the-art methods on benchmark datasets. The top three results are marked in \textcolor{red}{red}, \textcolor{blue}{blue} and \textcolor{green}{green}. ``$\dagger$'' indicates that methods adopt pre-training strategy on ImageNet.}
\vspace{-0.2cm}
\label{quantitative results}
\resizebox{2.1\columnwidth}{!}{
\begin{tabular}{|l|c|c|c|c|c|c|c|c|c|c|c|c|}
\hline
\multirow{2}{*}{Method} & \multirow{2}{*}{Scale} & \multirow{2}{*}{\makecell{Training\\Dataset}} &
\multicolumn{2}{c|}{Set5} &  \multicolumn{2}{c|}{Set14} &  \multicolumn{2}{c|}{BSD100} &  \multicolumn{2}{c|}{Urban100} &  \multicolumn{2}{c|}{Manga109}  
\\ 
\cline{4-13}
&  &  & PSNR & SSIM & PSNR & SSIM & PSNR & SSIM & PSNR & SSIM & PSNR & SSIM 
\\ 
\hline
\hline
EDSR & $\times$2 & DIV2K %
& 38.11
& 0.9602
& 33.92
& 0.9195
& 32.32
& 0.9013
& 32.93
& 0.9351
& 39.10
& 0.9773
\\
RCAN & $\times$2 & DIV2K %
& 38.27
& 0.9614
& 34.12
& 0.9216
& 32.41
& 0.9027
& 33.34
& 0.9384
& 39.44
& 0.9786
\\  
SAN & $\times$2 & DIV2K %
& {38.31}
& {0.9620}
& {34.07}
& {0.9213}
& {32.42}
& {0.9028}
& {33.10}
& {0.9370}
& {39.32}
& {0.9792}\\
IGNN & $\times$2 & DIV2K %
& {38.24}
& {0.9613}
& {34.07}
& {0.9217}
& {32.41}
& {0.9025}
& {33.23}
& {0.9383}
& {39.35}
& {0.9786}
\\
HAN & $\times$2 & DIV2K %
& {38.27}
& {0.9614}
& {34.16}
& {0.9217}
& {32.41}
& {0.9027}
& {33.35}
& {0.9385}
& {39.46}
& {0.9785}              
\\
NLSN & $\times$2 & DIV2K %
& 38.34 
& 0.9618 
& 34.08 
& 0.9231
& 32.43 
& 0.9027 
& 33.42
& 0.9394
& 39.59
& 0.9789
\\
RCAN-it & $\times$2 & DF2K %
& 38.37
& 0.9620
& 34.49
& 0.9250
& 32.48
& 0.9034
& 33.62
& 0.9410
& 39.88
& 0.9799
\\
SwinIR & $\times$2 & DF2K %
& 38.42
& 0.9623
& 34.46
& 0.9250
& 32.53
& 0.9041
& 33.81
& 0.9427
& 39.92
& 0.9797
\\
EDT & $\times$2 & DF2K %
& 38.45
& 0.9624
& 34.57
& 0.9258
& 32.52
& 0.9041
& 33.80
& 0.9425
& 39.93
& 0.9800
\\
\textbf{HAT-S} (ours) & $\times$2 & DF2K %
& {38.58}
& {0.9628}
& {34.70}
& {0.9261}
& {32.59}
& {0.9050}
& {34.31}
& {0.9459}
& {40.14}
& {0.9805}
\\
\textbf{HAT} (ours) & $\times$2 & DF2K %
& \textcolor{green}{38.63}
& {0.9630}
& \textcolor{green}{34.86}
& \textcolor{green}{0.9274}
& \textcolor{green}{32.62}
& \textcolor{green}{0.9053}
& \textcolor{green}{34.45}
& \textcolor{green}{0.9466}
& {40.26}
& {0.9809}
\\
\hdashline
IPT$^\dagger$ & $\times$2 & ImageNet %
& {38.37}
& {-}
& {34.43}
& {-}
& {32.48}
& {-}
& {33.76}
& {-}
& {-}
& {-}
\\
EDT$^\dagger$ & $\times$2 & DF2K %
& \textcolor{green}{38.63}
& \textcolor{green}{0.9632}
& 34.80
& 0.9273
& \textcolor{green}{32.62}
& {0.9052}
& 34.27
& 0.9456
& \textcolor{green}{40.37}
& \textcolor{green}{0.9811}
\\
\textbf{HAT}$^\dagger$ (ours) & $\times$2 & DF2K %
& \textcolor{blue}{38.73}
& \textcolor{blue}{0.9637}
& \textcolor{blue}{35.13}
& \textcolor{blue}{0.9282}
& \textcolor{blue}{32.69}
& \textcolor{blue}{0.9060}
& \textcolor{blue}{34.81}
& \textcolor{blue}{0.9489}
& \textcolor{blue}{40.71}
& \textcolor{blue}{0.9819}
\\
\textbf{HAT-L}$^\dagger$ (ours) & $\times$2 & DF2K %
& \textcolor{red}{38.91}
& \textcolor{red}{0.9646}
& \textcolor{red}{35.29}
& \textcolor{red}{0.9293}
& \textcolor{red}{32.74}
& \textcolor{red}{0.9066}
& \textcolor{red}{35.09}
& \textcolor{red}{0.9505}
& \textcolor{red}{41.01}
& \textcolor{red}{0.9831}
\\
\hline
\hline
EDSR & $\times$3 & DIV2K %
& 34.65
& 0.9280
& 30.52
& 0.8462
& 29.25
& 0.8093
& 28.80
& 0.8653
& 34.17
& 0.9476
\\
RCAN & $\times$3 & DIV2K %
& 34.74
& 0.9299
& 30.65
& 0.8482
& 29.32
& 0.8111
& 29.09
& 0.8702
& 34.44
& 0.9499
\\
SAN & $\times$3 & DIV2K %
& {34.75}
& {0.9300}
& {30.59}
& {0.8476}
& {29.33}
& {0.8112}
& {28.93}
& {0.8671}
& {34.30}
& {0.9494}
\\
IGNN & $\times$3 & DIV2K %
& {34.72}
& {0.9298}
& {30.66}
& {0.8484}
& {29.31}
& {0.8105}
& {29.03}
& {0.8696}
& {34.39}
& {0.9496}
\\
HAN  & $\times$3 & DIV2K %
& {34.75}
& {0.9299}
& {30.67}
& {0.8483}
& {29.32}
& {0.8110}
& {29.10}
& {0.8705}
& {34.48}
& {0.9500}
\\
NLSN & $\times$3 & DIV2K %
& 34.85 
& 0.9306 
& 30.70 
& 0.8485 
& 29.34 
& 0.8117 
& {29.25}
& {0.8726}
& 34.57 
& 0.9508  
\\
RCAN-it & $\times$3 & DF2K %
& {34.86}
& {0.9308}
& {30.76}
& {0.8505}
& {29.39}
& {0.8125}
& {29.38}
& {0.8755}
& {34.92}
& {0.9520}
\\
SwinIR & $\times$3 & DF2K %
& 34.97
& 0.9318
& 30.93
& 0.8534
& 29.46
& 0.8145
& 29.75
& 0.8826
& 35.12
& 0.9537
\\
EDT & $\times$3 & DF2K %
& 34.97
& 0.9316
& 30.89
& 0.8527
& 29.44
& 0.8142
& 29.72
& 0.8814
& 35.13
& 0.9534
\\
\textbf{HAT-S} (ours) & $\times$3 & DF2K %
& {35.01}
& {0.9325}
& {31.05}
& {0.8550}
& {29.50}
& {0.8158}
& {30.15}
& {0.8879}
& {35.40}
& {0.9547}
\\
\textbf{HAT} (ours) & $\times$3 & DF2K %
& {35.07}
& \textcolor{green}{0.9329}
& {31.08}
& \textcolor{green}{0.8555}
& \textcolor{green}{29.54}
& \textcolor{green}{0.8167}
& \textcolor{green}{30.23}
& \textcolor{green}{0.8896}
& \textcolor{green}{35.53}
& \textcolor{green}{0.9552}
\\
\hdashline
IPT$^\dagger$ & $\times$3 & ImageNet %
& {34.81}
& {-}
& {30.85}
& {-}
& {29.38}
& {-}
& {29.49}
& {-}
& {-}
& {-}
\\
EDT$^\dagger$ & $\times$3 & DF2K %
& \textcolor{green}{35.13}
& 0.9328
& \textcolor{green}{31.09}
& 0.8553
& {29.53}
& {0.8165}
& 30.07
& 0.8863
& 35.47
& 0.9550
\\
\textbf{HAT}$^\dagger$ (ours) & $\times$3 & DF2K %
& \textcolor{blue}{35.16}
& \textcolor{blue}{0.9335}
& \textcolor{blue}{31.33}
& \textcolor{blue}{0.8576}
& \textcolor{blue}{29.59}
& \textcolor{blue}{0.8177}
& \textcolor{blue}{30.70}
& \textcolor{blue}{0.8949}
& \textcolor{blue}{35.84}
& \textcolor{blue}{0.9567}
\\
\textbf{HAT-L}$^\dagger$ (ours) & $\times$3 & DF2K %
& \textcolor{red}{35.28}
& \textcolor{red}{0.9345}
& \textcolor{red}{31.47}
& \textcolor{red}{0.8584}
& \textcolor{red}{29.63}
& \textcolor{red}{0.8191}
& \textcolor{red}{30.92}
& \textcolor{red}{0.8981}
& \textcolor{red}{36.02}
& \textcolor{red}{0.9576}
\\
\hline
\hline
EDSR & $\times$4 & DIV2K %
& 32.46
& 0.8968
& 28.80
& 0.7876
& 27.71
& 0.7420
& 26.64
& 0.8033
& 31.02
& 0.9148
\\
RCAN & $\times$4 & DIV2K %
& 32.63
& 0.9002
& 28.87
& 0.7889
& 27.77
& 0.7436
& 26.82
& 0.8087
& 31.22
& 0.9173
\\
SAN & $\times$4 & DIV2K %
& {32.64}
& {0.9003}
& {28.92}
& {0.7888}
& {27.78}
& {0.7436}
& {26.79}
& {0.8068}
& {31.18}
& {0.9169}
\\
IGNN & $\times$4 & DIV2K %
& {32.57}
& {0.8998}
& {28.85}
& {0.7891}
& {27.77}
& {0.7434}
& {26.84}
& {0.8090}
& {31.28}
& {0.9182}
\\
HAN & $\times$4 & DIV2K %
& {32.64}
& {0.9002}
& {28.90}
& {0.7890}
& {27.80}
& {0.7442}
& {26.85}
& {0.8094}
& {31.42}
& {0.9177}
\\
NLSN & $\times$4 & DIV2K %
& 32.59 
& 0.9000 
& 28.87 
& 0.7891 
& 27.78 
& 0.7444 
& {26.96}
& {0.8109}
& 31.27 
& 0.9184
\\
RRDB & $\times$4 & DF2K %
& {32.73}
& {0.9011 }
& {28.99}
& {0.7917}
& {27.85}
& {0.7455}
& {27.03}
& {0.8153}
& {31.66}
& {0.9196}
\\
RCAN-it & $\times$4 & DF2K %
& 32.69
& 0.9007
& 28.99
& 0.7922
& 27.87
& 0.7459
& 27.16
& 0.8168
& 31.78
& 0.9217
\\
SwinIR & $\times$4 & DF2K %
& 32.92
& 0.9044
& 29.09
& 0.7950
& 27.92
& 0.7489
& 27.45
& 0.8254
& 32.03
& 0.9260
\\
EDT & $\times$4 & DF2K %
& 32.82
& 0.9031
& 29.09
& 0.7939
& 27.91
& 0.7483
& 27.46
& 0.8246
& 32.05
& 0.9254
\\
\textbf{HAT-S} (ours) & $\times$4 & DF2K %
& {32.92}
& {0.9047}
& {29.15}
& {0.7958}
& {27.97}
& {0.7505}
& {27.87}
& {0.8346}
& {32.35}
& {0.9283}
\\
\textbf{HAT} (ours) & $\times$4 & DF2K %
& {33.04}
& \textcolor{green}{0.9056}
& \textcolor{green}{29.23}
& \textcolor{green}{0.7973}
& \textcolor{green}{28.00}
& \textcolor{green}{0.7517}
& \textcolor{green}{27.97}
& \textcolor{green}{0.8368}
& \textcolor{green}{32.48}
& \textcolor{green}{0.9292}
\\
\hdashline
IPT$^\dagger$ & $\times$4 & ImageNet %
& {32.64}
& {-}
& {29.01}
& {-}
& {27.82}
& {-}
& {27.26}
& {-}
& {-}
& {-}
\\
EDT$^\dagger$ & $\times$4 & DF2K %
& \textcolor{green}{33.06}
& {0.9055}
& \textcolor{green}{29.23}
& {0.7971}
& {27.99}
& 0.7510
& 27.75
& 0.8317
& 32.39
& 0.9283
\\
\textbf{HAT}$^\dagger$ (ours) & $\times$4 & DF2K %
& \textcolor{blue}{33.18}
& \textcolor{blue}{0.9073}
& \textcolor{blue}{29.38}
& \textcolor{blue}{0.8001}
& \textcolor{blue}{28.05}
& \textcolor{blue}{0.7534}
& \textcolor{blue}{28.37}
& \textcolor{blue}{0.8447}
& \textcolor{blue}{32.87}
& \textcolor{blue}{0.9319}
\\
\textbf{HAT-L}$^\dagger$ (ours) & $\times$4 & DF2K %
& \textcolor{red}{33.30}
& \textcolor{red}{0.9083}
& \textcolor{red}{29.47}
& \textcolor{red}{0.8015}
& \textcolor{red}{28.09}
& \textcolor{red}{0.7551}
& \textcolor{red}{28.60}
& \textcolor{red}{0.8498}
& \textcolor{red}{33.09}
& \textcolor{red}{0.9335}
\\
\hline             
\end{tabular}
}
\vspace{-0.5cm}
\end{center}
\end{table*}

\subsection{Comparison with State-of-the-Art Methods}

\noindent
\textbf{Quantitative results.}
Tab.~\ref{quantitative results} shows the quantitative comparison of our approach and the state-of-the-art methods: EDSR~\cite{edsr}, RCAN~\cite{rcan}, SAN~\cite{san}, IGNN \cite{ignn}, HAN~\cite{han}, NLSN~\cite{nlsn}, RCAN-it~\cite{rcanit}, as well as approaches using ImageNet pre-training, \textit{i.e.}, IPT~\cite{ipt} and EDT~\cite{edt}. We can see that our method outperforms the other methods significantly on all benchmark datasets. Concretely, HAT surpasses SwinIR by 0.48dB$\sim$0.64dB on Urban100 and 0.34dB$\sim$0.45dB on Manga109. When compared with the approaches using pre-training, HAT also has large performance gains of more than 0.5dB against EDT on Urban100 for all three scales. Besides, HAT with pre-training outperforms SwinIR by a huge margin of up to 1dB on Urban100 for $\times$2 SR. Moreover, the large model HAT-L can even bring further improvement and greatly expands the performance upper bound of this task. HAT-S with fewer parameters and similar computation can also significantly outperforms the state-of-the-art method SwinIR. (Detailed computational complexity comparison can be found in the \textit{supp.} file.) Note that the performance gaps are much larger on Urban100, as it contains more structured and self-repeated patterns that can provide more useful pixels for reconstruction when the utilized range of information is enlarged. All these results show the effectiveness of our method. 

\begin{figure*}[!t]
\begin{center}
\includegraphics[width=1.0\textwidth]{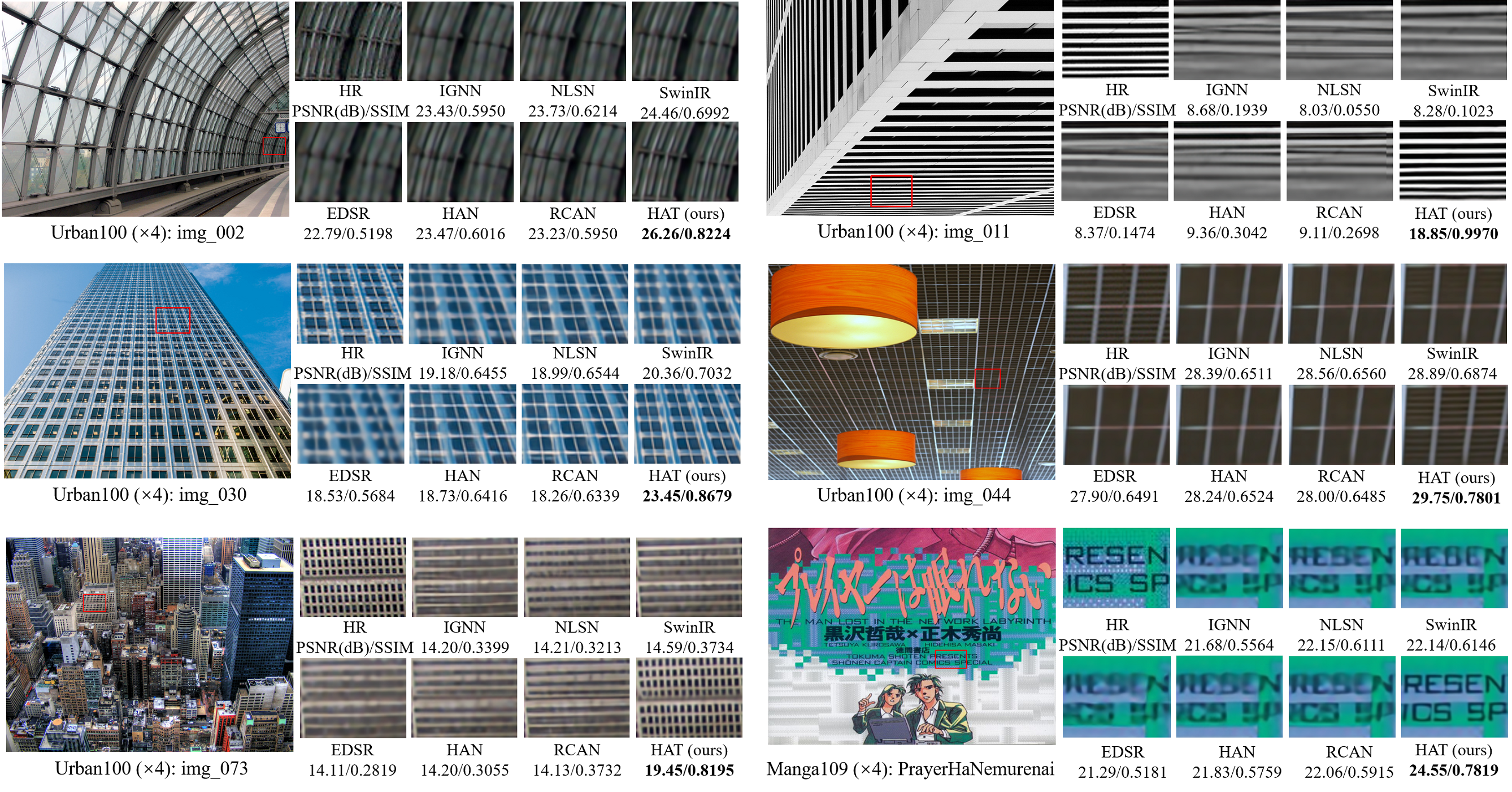}
\end{center}
\vspace{-0.5cm}
\caption{Visual comparison on $\times$4 SR. The patches for comparison are marked with red boxes in the original images. PSNR/SSIM is calculated based on the patches to better reflect the performance difference.}
\label{visual_cmp}
\vspace{-0.5cm}
\end{figure*}

\noindent
\textbf{Visual comparison.}
We provide the visual comparison in Fig.~\ref{visual_cmp}. For the images ``img\underline{~}002'', ``img\underline{~}011'', ``img\underline{~}030'', ``img\underline{~}044'' and ``img\underline{~}073'' in Urban100, HAT successfully recovers the clear lattice content. In contrast, the other approaches all suffer from severe blurry effects. We can also observe similar behaviors on ``PrayerHaNemurenai'' in Manga109. When recovering the characters, HAT obtains significantly clearer textures than other methods. The visual results also demonstrate the superiority of our approach.

\subsection{Study on the pre-training strategy}
\label{cmp_pretrain}
\begin{figure}[!t]
\begin{center}
\includegraphics[width=1\linewidth]{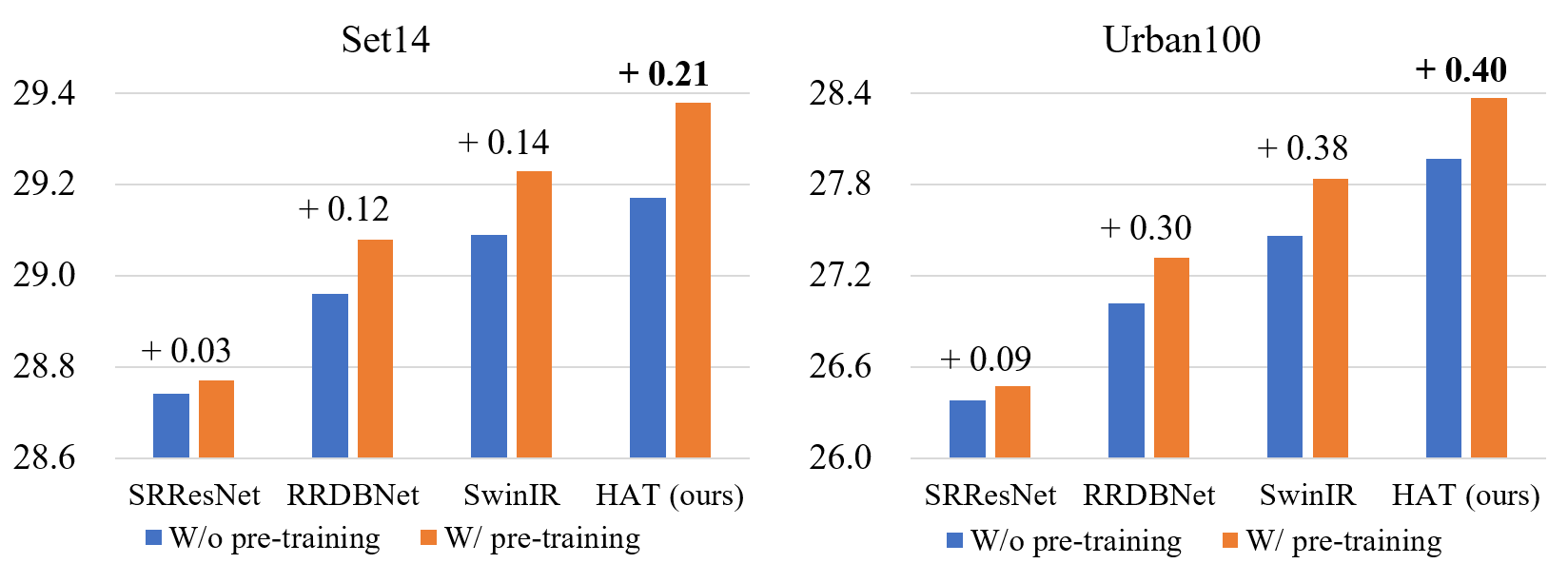}
\end{center}
\vspace{-0.4cm}
\caption{Quantitative comparison on PSNR(dB) of four different networks without and with the same-task pre-training on $\times$4 SR.}
\label{pretrain2}
\vspace{-0.5cm}
\end{figure}

In Tab.~\ref{quantitative results}, we can see that HAT can benefit greatly from the pre-training strategy, by comparing the performance of HAT and {HAT}$^\dagger$. To show the superiority of the proposed same-task pre-training, we also apply the multi-related-task pre-training~\cite{edt} to HAT for comparison using full ImageNet, under the same training settings as ~\cite{edt}. As depicted as Tab.~\ref{pretrain1}, the same-task pre-training performs better, not only in the pre-training stage but also in the fine-tuning process. From this perspective, multi-task pre-training probably impairs the restoration performance of the network on a specific degradation, while the same-task pre-training can maximize the performance gain brought by large-scale data. To further investigate the influences of our pre-training strategy for different networks, we apply our pre-training to four networks: SRResNet (1.5M), RRDBNet (16.7M), SwinIR (11.9M) and HAT (20.8M), as shown in Fig.~\ref{pretrain2}. First, we can see that all four networks can benefit from pre-training, showing the effectiveness of the proposed same-task pre-training strategy. Second, for the same type of network (\textit{i.e.}, CNN or Transformer), the larger the network capacity, the more performance gain from pre-training. Third, although with less parameters, SwinIR obtains greater performance improvement from the pre-training compared to RRDBNet. It suggests that Transformer needs more data to exploit the potential of the model. Finally, HAT obtains the largest gain from pre-training, indicating the necessity of the pre-training strategy for such large models. Equipped with big models and large-scale data, we show the performance upper bound of this task is significantly extended.


\section{Conclusion}
In this paper, we propose a novel Hybrid Attention Transformer, HAT, for single image super-resolution. Our model combines channel attention and self-attention to activate more pixels for high-resolution reconstruction. Besides, we propose an overlapping cross-attention module to enhance the interaction of cross-window information. Moreover, we introduce a same-task pre-training strategy to further exploit the potential of HAT. Extensive experiments show the effectiveness of the proposed modules and the pre-training strategy. Our approach significantly outperforms the state-of-the-art methods quantitatively and qualitatively. 

\begin{table}[!t]
\center
\begin{center}
\caption{Quantitative results on PSNR(dB) of HAT using two kinds of pre-training strategies on $\times$4 SR under the same training setting. The full ImageNet dataset is adopted to perform pre-training and DF2K dataset is used for fine-tuning.}
\vspace{-0.2cm}
\label{pretrain1}
\setlength{\tabcolsep}{1.3mm}{
\begin{tabular}{c|c|ccc} 
\hline 
Strategy & Stage & Set5 & Set14 & Urban100 \\
\hline 
Multi-related-task & pre-training & 32.94 & 29.17 & 28.05\\
pre-training & fine-tuning & 33.06 & 29.33 & 28.21\\
\hline 
Same-task & pre-training & 33.02 & 29.20 & 28.11\\
pre-training(ours) & fine-tuning & 33.07 & 29.34 & 28.28\\
\hline 
\end{tabular}
}
\end{center}
\end{table}

\noindent\textbf{Acknowledgement.} This work was supported in part by Macau Science and Technology Development Fund under 

\noindent SKLIOTSC-2021-2023, 0072/2020/AMJ, 0022/2022/A1; 

\noindent in part by Alibaba Innovative Research Program; in part by the National Natural Science Foundation of China under Grant (61971476, 62276251), the Joint Lab of CAS-HK; and in part by the Youth Innovation Promotion Association of Chinese Academy of Sciences (No. 2020356).

{\small
\bibliographystyle{ieee_fullname}
\bibliography{cvpr23}

\begin{thebibliography}{10}\itemsep=-1pt

\bibitem{bao2021beit}
Hangbo Bao, Li Dong, and Furu Wei.
\newblock Beit: Bert pre-training of image transformers.
\newblock {\em arXiv preprint arXiv:2106.08254}, 2021.

\bibitem{set5}
Marco Bevilacqua, Aline Roumy, Christine Guillemot, and Marie~Line
  Alberi-Morel.
\newblock Low-complexity single-image super-resolution based on nonnegative
  neighbor embedding.
\newblock 2012.

\bibitem{cao2021swin}
Hu Cao, Yueyue Wang, Joy Chen, Dongsheng Jiang, Xiaopeng Zhang, Qi Tian, and
  Manning Wang.
\newblock Swin-unet: Unet-like pure transformer for medical image segmentation,
  2021.

\bibitem{vsrt}
Jiezhang Cao, Yawei Li, Kai Zhang, and Luc Van~Gool.
\newblock Video super-resolution transformer, 2021.

\bibitem{detr}
Nicolas Carion, Francisco Massa, Gabriel Synnaeve, Nicolas Usunier, Alexander
  Kirillov, and Sergey Zagoruyko.
\newblock End-to-end object detection with transformers.
\newblock In {\em European conference on computer vision}, pages 213--229.
  Springer, 2020.

\bibitem{ipt}
Hanting Chen, Yunhe Wang, Tianyu Guo, Chang Xu, Yiping Deng, Zhenhua Liu, Siwei
  Ma, Chunjing Xu, Chao Xu, and Wen Gao.
\newblock Pre-trained image processing transformer.
\newblock In {\em Proceedings of the IEEE/CVF Conference on Computer Vision and
  Pattern Recognition}, pages 12299--12310, 2021.

\bibitem{twins}
Xiangxiang Chu, Zhi Tian, Yuqing Wang, Bo Zhang, Haibing Ren, Xiaolin Wei,
  Huaxia Xia, and Chunhua Shen.
\newblock Twins: Revisiting the design of spatial attention in vision
  transformers.
\newblock {\em Advances in Neural Information Processing Systems}, 34, 2021.

\bibitem{san}
Tao Dai, Jianrui Cai, Yongbing Zhang, Shu-Tao Xia, and Lei Zhang.
\newblock Second-order attention network for single image super-resolution.
\newblock In {\em Proceedings of the IEEE/CVF conference on computer vision and
  pattern recognition}, pages 11065--11074, 2019.

\bibitem{imagenet}
Jia Deng, Wei Dong, Richard Socher, Li-Jia Li, Kai Li, and Li Fei-Fei.
\newblock Imagenet: A large-scale hierarchical image database.
\newblock In {\em 2009 IEEE conference on computer vision and pattern
  recognition}, pages 248--255, 2009.

\bibitem{srcnn_eccv}
Chao Dong, Chen~Change Loy, Kaiming He, and Xiaoou Tang.
\newblock Learning a deep convolutional network for image super-resolution.
\newblock In {\em European conference on computer vision}, pages 184--199.
  Springer, 2014.

\bibitem{srcnn_tpami}
Chao Dong, Chen~Change Loy, Kaiming He, and Xiaoou Tang.
\newblock Image super-resolution using deep convolutional networks.
\newblock {\em IEEE transactions on pattern analysis and machine intelligence},
  38(2):295--307, 2015.

\bibitem{fsrcnn}
Chao Dong, Chen~Change Loy, and Xiaoou Tang.
\newblock Accelerating the super-resolution convolutional neural network.
\newblock In {\em European conference on computer vision}, pages 391--407.
  Springer, 2016.

\bibitem{cswin}
Xiaoyi Dong, Jianmin Bao, Dongdong Chen, Weiming Zhang, Nenghai Yu, Lu Yuan,
  Dong Chen, and Baining Guo.
\newblock Cswin transformer: A general vision transformer backbone with
  cross-shaped windows.
\newblock In {\em Proceedings of the IEEE/CVF Conference on Computer Vision and
  Pattern Recognition}, pages 12124--12134, 2022.

\bibitem{vit}
Alexey Dosovitskiy, Lucas Beyer, Alexander Kolesnikov, Dirk Weissenborn,
  Xiaohua Zhai, Thomas Unterthiner, Mostafa Dehghani, Matthias Minderer, Georg
  Heigold, Sylvain Gelly, et~al.
\newblock An image is worth 16x16 words: Transformers for image recognition at
  scale, 2020.

\bibitem{lam}
Jinjin Gu and Chao Dong.
\newblock Interpreting super-resolution networks with local attribution maps.
\newblock In {\em Proceedings of the IEEE/CVF Conference on Computer Vision and
  Pattern Recognition}, pages 9199--9208, 2021.

\bibitem{he2022masked}
Kaiming He, Xinlei Chen, Saining Xie, Yanghao Li, Piotr Doll{\'a}r, and Ross
  Girshick.
\newblock Masked autoencoders are scalable vision learners.
\newblock In {\em Proceedings of the IEEE/CVF Conference on Computer Vision and
  Pattern Recognition}, pages 16000--16009, 2022.

\bibitem{GELU}
Dan Hendrycks and Kevin Gimpel.
\newblock Gaussian error linear units (gelus), 2016.

\bibitem{dat}
Gao Huang, Yulin Wang, Kangchen Lv, Haojun Jiang, Wenhui Huang, Pengfei Qi, and
  Shiji Song.
\newblock Glance and focus networks for dynamic visual recognition, 2022.

\bibitem{urban100}
Jia-Bin Huang, Abhishek Singh, and Narendra Ahuja.
\newblock Single image super-resolution from transformed self-exemplars.
\newblock In {\em Proceedings of the IEEE conference on computer vision and
  pattern recognition}, pages 5197--5206, 2015.

\bibitem{shufflet}
Zilong Huang, Youcheng Ben, Guozhong Luo, Pei Cheng, Gang Yu, and Bin Fu.
\newblock Shuffle transformer: Rethinking spatial shuffle for vision
  transformer, 2021.

\bibitem{vdsr}
Jiwon Kim, Jung~Kwon Lee, and Kyoung~Mu Lee.
\newblock Accurate image super-resolution using very deep convolutional
  networks.
\newblock In {\em Proceedings of the IEEE conference on computer vision and
  pattern recognition}, pages 1646--1654, 2016.

\bibitem{drcn}
Jiwon Kim, Jung~Kwon Lee, and Kyoung~Mu Lee.
\newblock Deeply-recursive convolutional network for image super-resolution.
\newblock In {\em Proceedings of the IEEE conference on computer vision and
  pattern recognition}, pages 1637--1645, 2016.

\bibitem{dropoutSR}
Xiangtao Kong, Xina Liu, Jinjin Gu, Yu Qiao, and Chao Dong.
\newblock Reflash dropout in image super-resolution.
\newblock In {\em Proceedings of the IEEE/CVF Conference on Computer Vision and
  Pattern Recognition}, pages 6002--6012, 2022.

\bibitem{classsr}
Xiangtao Kong, Hengyuan Zhao, Yu Qiao, and Chao Dong.
\newblock Classsr: A general framework to accelerate super-resolution networks
  by data characteristic.
\newblock In {\em Proceedings of the IEEE/CVF Conference on Computer Vision and
  Pattern Recognition (CVPR)}, pages 12016--12025, June 2021.

\bibitem{srgan}
Christian Ledig, Lucas Theis, Ferenc Husz{\'a}r, Jose Caballero, Andrew
  Cunningham, Alejandro Acosta, Andrew Aitken, Alykhan Tejani, Johannes Totz,
  Zehan Wang, et~al.
\newblock Photo-realistic single image super-resolution using a generative
  adversarial network.
\newblock In {\em Proceedings of the IEEE conference on computer vision and
  pattern recognition}, pages 4681--4690, 2017.

\bibitem{uniformer}
Kunchang Li, Yali Wang, Junhao Zhang, Peng Gao, Guanglu Song, Yu Liu, Hongsheng
  Li, and Yu Qiao.
\newblock Uniformer: Unifying convolution and self-attention for visual
  recognition, 2022.

\bibitem{edt}
Wenbo Li, Xin Lu, Jiangbo Lu, Xiangyu Zhang, and Jiaya Jia.
\newblock On efficient transformer and image pre-training for low-level vision,
  2021.

\bibitem{localvit}
Yawei Li, Kai Zhang, Jiezhang Cao, Radu Timofte, and Luc Van~Gool.
\newblock Localvit: Bringing locality to vision transformers, 2021.

\bibitem{bsrn}
Zheyuan Li, Yingqi Liu, Xiangyu Chen, Haoming Cai, Jinjin Gu, Yu Qiao, and Chao
  Dong.
\newblock Blueprint separable residual network for efficient image
  super-resolution.
\newblock In {\em Proceedings of the IEEE/CVF Conference on Computer Vision and
  Pattern Recognition (CVPR) Workshops}, pages 833--843, June 2022.

\bibitem{vrt}
Jingyun Liang, Jiezhang Cao, Yuchen Fan, Kai Zhang, Rakesh Ranjan, Yawei Li,
  Radu Timofte, and Luc Van~Gool.
\newblock Vrt: A video restoration transformer, 2022.

\bibitem{swinir}
Jingyun Liang, Jiezhang Cao, Guolei Sun, Kai Zhang, Luc Van~Gool, and Radu
  Timofte.
\newblock Swinir: Image restoration using swin transformer.
\newblock In {\em Proceedings of the IEEE/CVF International Conference on
  Computer Vision}, pages 1833--1844, 2021.

\bibitem{edsr}
Bee Lim, Sanghyun Son, Heewon Kim, Seungjun Nah, and Kyoung Mu~Lee.
\newblock Enhanced deep residual networks for single image super-resolution.
\newblock In {\em Proceedings of the IEEE conference on computer vision and
  pattern recognition workshops}, pages 136--144, 2017.

\bibitem{div2k}
Bee Lim, Sanghyun Son, Heewon Kim, Seungjun Nah, and Kyoung Mu~Lee.
\newblock Enhanced deep residual networks for single image super-resolution.
\newblock In {\em Proceedings of the IEEE conference on computer vision and
  pattern recognition workshops}, pages 136--144, 2017.

\bibitem{rcanit}
Zudi Lin, Prateek Garg, Atmadeep Banerjee, Salma~Abdel Magid, Deqing Sun, Yulun
  Zhang, Luc Van~Gool, Donglai Wei, and Hanspeter Pfister.
\newblock Revisiting rcan: Improved training for image super-resolution, 2022.

\bibitem{nlrn}
Ding Liu, Bihan Wen, Yuchen Fan, Chen~Change Loy, and Thomas~S Huang.
\newblock Non-local recurrent network for image restoration.
\newblock {\em Advances in neural information processing systems}, 31, 2018.

\bibitem{liu2020deep}
Li Liu, Wanli Ouyang, Xiaogang Wang, Paul Fieguth, Jie Chen, Xinwang Liu, and
  Matti Pietik{\"a}inen.
\newblock Deep learning for generic object detection: A survey.
\newblock {\em International journal of computer vision}, 128(2):261--318,
  2020.

\bibitem{ddr}
Yihao Liu, Anran Liu, Jinjin Gu, Zhipeng Zhang, Wenhao Wu, Yu Qiao, and Chao
  Dong.
\newblock Discovering" semantics" in super-resolution networks, 2021.

\bibitem{srga}
Yihao Liu, Hengyuan Zhao, Jinjin Gu, Yu Qiao, and Chao Dong.
\newblock Evaluating the generalization ability of super-resolution networks.
\newblock {\em arXiv preprint arXiv:2205.07019}, 2022.

\bibitem{swin_t}
Ze Liu, Yutong Lin, Yue Cao, Han Hu, Yixuan Wei, Zheng Zhang, Stephen Lin, and
  Baining Guo.
\newblock Swin transformer: Hierarchical vision transformer using shifted
  windows.
\newblock In {\em Proceedings of the IEEE/CVF International Conference on
  Computer Vision}, pages 10012--10022, 2021.

\bibitem{bsd100}
David Martin, Charless Fowlkes, Doron Tal, and Jitendra Malik.
\newblock A database of human segmented natural images and its application to
  evaluating segmentation algorithms and measuring ecological statistics.
\newblock In {\em Proceedings Eighth IEEE International Conference on Computer
  Vision. ICCV 2001}, volume~2, pages 416--423. IEEE, 2001.

\bibitem{manga109}
Yusuke Matsui, Kota Ito, Yuji Aramaki, Azuma Fujimoto, Toru Ogawa, Toshihiko
  Yamasaki, and Kiyoharu Aizawa.
\newblock Sketch-based manga retrieval using manga109 dataset.
\newblock {\em Multimedia Tools and Applications}, 76(20):21811--21838, 2017.

\bibitem{nlsn}
Yiqun Mei, Yuchen Fan, and Yuqian Zhou.
\newblock Image super-resolution with non-local sparse attention.
\newblock In {\em Proceedings of the IEEE/CVF Conference on Computer Vision and
  Pattern Recognition}, pages 3517--3526, 2021.

\bibitem{han}
Ben Niu, Weilei Wen, Wenqi Ren, Xiangde Zhang, Lianping Yang, Shuzhen Wang,
  Kaihao Zhang, Xiaochun Cao, and Haifeng Shen.
\newblock Single image super-resolution via a holistic attention network.
\newblock In {\em European conference on computer vision}, pages 191--207.
  Springer, 2020.

\bibitem{moa}
Krushi Patel, Andres~M Bur, Fengjun Li, and Guanghui Wang.
\newblock Aggregating global features into local vision transformer, 2022.

\bibitem{cka}
Maithra Raghu, Thomas Unterthiner, Simon Kornblith, Chiyuan Zhang, and Alexey
  Dosovitskiy.
\newblock Do vision transformers see like convolutional neural networks?
\newblock {\em Advances in Neural Information Processing Systems}, 34, 2021.

\bibitem{ramachandran2019studying}
Prajit Ramachandran, Niki Parmar, Ashish Vaswani, Irwan Bello, Anselm Levskaya,
  and Jon Shlens.
\newblock Studying stand-alone self-attention in vision models.
\newblock 2019.

\bibitem{pixelshuffle}
Wenzhe Shi, Jose Caballero, Ferenc Husz{\'a}r, Johannes Totz, Andrew~P Aitken,
  Rob Bishop, Daniel Rueckert, and Zehan Wang.
\newblock Real-time single image and video super-resolution using an efficient
  sub-pixel convolutional neural network.
\newblock In {\em Proceedings of the IEEE conference on computer vision and
  pattern recognition}, pages 1874--1883, 2016.

\bibitem{drrn}
Ying Tai, Jian Yang, and Xiaoming Liu.
\newblock Image super-resolution via deep recursive residual network.
\newblock In {\em Proceedings of the IEEE conference on computer vision and
  pattern recognition}, pages 3147--3155, 2017.

\bibitem{flicker2k}
Radu Timofte, Eirikur Agustsson, Luc Van~Gool, Ming-Hsuan Yang, and Lei Zhang.
\newblock Ntire 2017 challenge on single image super-resolution: Methods and
  results.
\newblock In {\em Proceedings of the IEEE conference on computer vision and
  pattern recognition workshops}, pages 114--125, 2017.

\bibitem{touvron2021training}
Hugo Touvron, Matthieu Cord, Matthijs Douze, Francisco Massa, Alexandre
  Sablayrolles, and Herv{\'e} J{\'e}gou.
\newblock Training data-efficient image transformers \& distillation through
  attention.
\newblock In {\em International Conference on Machine Learning}, pages
  10347--10357. PMLR, 2021.

\bibitem{maxim}
Zhengzhong Tu, Hossein Talebi, Han Zhang, Feng Yang, Peyman Milanfar, Alan
  Bovik, and Yinxiao Li.
\newblock Maxim: Multi-axis mlp for image processing.
\newblock {\em CVPR}, 2022.

\bibitem{vaswani2021scaling}
Ashish Vaswani, Prajit Ramachandran, Aravind Srinivas, Niki Parmar, Blake
  Hechtman, and Jonathon Shlens.
\newblock Scaling local self-attention for parameter efficient visual
  backbones.
\newblock In {\em Proceedings of the IEEE/CVF Conference on Computer Vision and
  Pattern Recognition}, pages 12894--12904, 2021.

\bibitem{transformer}
Ashish Vaswani, Noam Shazeer, Niki Parmar, Jakob Uszkoreit, Llion Jones,
  Aidan~N Gomez, {\L}ukasz Kaiser, and Illia Polosukhin.
\newblock Attention is all you need.
\newblock {\em Advances in neural information processing systems}, 30, 2017.

\bibitem{pvt}
Wenhai Wang, Enze Xie, Xiang Li, Deng-Ping Fan, Kaitao Song, Ding Liang, Tong
  Lu, Ping Luo, and Ling Shao.
\newblock Pyramid vision transformer: A versatile backbone for dense prediction
  without convolutions.
\newblock In {\em Proceedings of the IEEE/CVF International Conference on
  Computer Vision}, pages 568--578, 2021.

\bibitem{realesrgan}
Xintao Wang, Liangbin Xie, Chao Dong, and Ying Shan.
\newblock Real-esrgan: Training real-world blind super-resolution with pure
  synthetic data.
\newblock In {\em Proceedings of the IEEE/CVF International Conference on
  Computer Vision}, pages 1905--1914, 2021.

\bibitem{esrgan}
Xintao Wang, Ke Yu, Shixiang Wu, Jinjin Gu, Yihao Liu, Chao Dong, Yu Qiao, and
  Chen Change~Loy.
\newblock Esrgan: Enhanced super-resolution generative adversarial networks.
\newblock In {\em Proceedings of the European conference on computer vision
  (ECCV) workshops}, pages 0--0, 2018.

\bibitem{uformer}
Zhendong Wang, Xiaodong Cun, Jianmin Bao, Wengang Zhou, Jianzhuang Liu, and
  Houqiang Li.
\newblock Uformer: A general u-shaped transformer for image restoration.
\newblock In {\em Proceedings of the IEEE/CVF Conference on Computer Vision and
  Pattern Recognition}, pages 17683--17693, 2022.

\bibitem{wu2020visual}
Bichen Wu, Chenfeng Xu, Xiaoliang Dai, Alvin Wan, Peizhao Zhang, Zhicheng Yan,
  Masayoshi Tomizuka, Joseph Gonzalez, Kurt Keutzer, and Peter Vajda.
\newblock Visual transformers: Token-based image representation and processing
  for computer vision, 2020.

\bibitem{cvt}
Haiping Wu, Bin Xiao, Noel Codella, Mengchen Liu, Xiyang Dai, Lu Yuan, and Lei
  Zhang.
\newblock Cvt: Introducing convolutions to vision transformers.
\newblock In {\em Proceedings of the IEEE/CVF International Conference on
  Computer Vision}, pages 22--31, 2021.

\bibitem{palet}
Sitong Wu, Tianyi Wu, Haoru Tan, and Guodong Guo.
\newblock Pale transformer: A general vision transformer backbone with
  pale-shaped attention.
\newblock In {\em Proceedings of the AAAI Conference on Artificial
  Intelligence}, volume~36, pages 2731--2739, 2022.

\bibitem{vitc}
Tete Xiao, Piotr Dollar, Mannat Singh, Eric Mintun, Trevor Darrell, and Ross
  Girshick.
\newblock Early convolutions help transformers see better.
\newblock {\em Advances in Neural Information Processing Systems}, 34, 2021.

\bibitem{faig}
Liangbin Xie, Xintao Wang, Chao Dong, Zhongang Qi, and Ying Shan.
\newblock Finding discriminative filters for specific degradations in blind
  super-resolution.
\newblock {\em Advances in Neural Information Processing Systems}, 34, 2021.

\bibitem{ceit}
Kun Yuan, Shaopeng Guo, Ziwei Liu, Aojun Zhou, Fengwei Yu, and Wei Wu.
\newblock Incorporating convolution designs into visual transformers.
\newblock In {\em Proceedings of the IEEE/CVF International Conference on
  Computer Vision}, pages 579--588, 2021.

\bibitem{hrformer}
Yuhui Yuan, Rao Fu, Lang Huang, Weihong Lin, Chao Zhang, Xilin Chen, and
  Jingdong Wang.
\newblock Hrformer: High-resolution vision transformer for dense predict.
\newblock {\em Advances in Neural Information Processing Systems},
  34:7281--7293, 2021.

\bibitem{restormer}
Syed~Waqas Zamir, Aditya Arora, Salman Khan, Munawar Hayat, Fahad~Shahbaz Khan,
  and Ming-Hsuan Yang.
\newblock Restormer: Efficient transformer for high-resolution image
  restoration.
\newblock In {\em Proceedings of the IEEE/CVF Conference on Computer Vision and
  Pattern Recognition}, pages 5728--5739, 2022.

\bibitem{set14}
Roman Zeyde, Michael Elad, and Matan Protter.
\newblock On single image scale-up using sparse-representations.
\newblock In {\em International conference on curves and surfaces}, pages
  711--730. Springer, 2010.

\bibitem{ranksrgan}
Wenlong Zhang, Yihao Liu, Chao Dong, and Yu Qiao.
\newblock Ranksrgan: Generative adversarial networks with ranker for image
  super-resolution.
\newblock In {\em Proceedings of the IEEE/CVF International Conference on
  Computer Vision}, pages 3096--3105, 2019.

\bibitem{rcan}
Yulun Zhang, Kunpeng Li, Kai Li, Lichen Wang, Bineng Zhong, and Yun Fu.
\newblock Image super-resolution using very deep residual channel attention
  networks.
\newblock In {\em Proceedings of the European conference on computer vision
  (ECCV)}, pages 286--301, 2018.

\bibitem{rnan}
Yulun Zhang, Kunpeng Li, Kai Li, Bineng Zhong, and Yun Fu.
\newblock Residual non-local attention networks for image restoration, 2019.

\bibitem{rdn}
Yulun Zhang, Yapeng Tian, Yu Kong, Bineng Zhong, and Yun Fu.
\newblock Residual dense network for image super-resolution.
\newblock In {\em Proceedings of the IEEE conference on computer vision and
  pattern recognition}, pages 2472--2481, 2018.

\bibitem{spach}
Yucheng Zhao, Guangting Wang, Chuanxin Tang, Chong Luo, Wenjun Zeng, and
  Zheng-Jun Zha.
\newblock A battle of network structures: An empirical study of cnn,
  transformer, and mlp, 2021.

\bibitem{ignn}
Shangchen Zhou, Jiawei Zhang, Wangmeng Zuo, and Chen~Change Loy.
\newblock Cross-scale internal graph neural network for image super-resolution.
\newblock {\em Advances in neural information processing systems},
  33:3499--3509, 2020.

\end{thebibliography}
}

\newpage
\clearpage
\appendix

\begin{minipage}{2\linewidth}
   \begin{center}
      \setlength\parindent{0pt}
      {\Large \bf  Activating More Pixels in Image Super-Resolution Transformer\\
      Supplementary Material\par}
            \vspace*{24pt}
      {
      \large
      \begin{tabular}[t]{c}
        Xiangyu Chen$^{1,2,3}$ \quad Xintao Wang$^{4}$ \quad
        Jiantao Zhou$^{1}$ \quad Yu Qiao$^{2,3}$ \quad Chao Dong$^{2,3}$\\[0.4em]
        $^1$State Key Laboratory of Internet of Things for Smart City, University of Macau\\
        $^2$Shenzhen Key Lab of Computer Vision and Pattern Recognition, \\Shenzhen Institute of Advanced Technology, Chinese Academy of Sciences\\
        $^3$Shanghai Artificial Intelligence Laboratory\quad
        $^4$ARC Lab, Tencent PCG\\
        [0.4em] \url{https://github.com/XPixelGroup/HAT}
        \vspace*{0.5cm}
      \end{tabular}
      \par
      }
   \end{center}
\end{minipage}

\section{Training Details}
We use DF2K (DIV2K+Flicker2K) with 3450 images as the training dataset when training from scratch. The low-resolution images are generated from the ground truth images by the ``bicubic'' down-sampling in MATLAB. We set the input patch size to $64\times 64$ and use random rotation and horizontally flipping for data augmentation. The mini-batch size is set to 32 and total training iterations are set to 500K. The learning rate is initialized as 2e-4 and reduced by half at [250K,400K,450K,475K]. For $\times$4 SR, we initialize the model with pre-trained $\times$2 SR weights and halve the iterations for each learning rate decay as well as total iterations. We adopt Adam optimizer with $\beta_1=0.9$ and $\beta_2=0.99$ to train the model. For the same-task pre-training, the full ImageNet dataset with 1.28 million images is first exploited to pre-train the model for 800K iterations. The initial learning rate is also set to 2e-4 but reduced by half at [300K,500K,650K,700K,750k]. Then, we adopt DF2K dataset to fine-tune the pre-trained model. For fine-tuning, we set the initial learning rate to 1e-5 and halve it at [125K,200K,230K,240K] for total 250K training iterations. 

\section{Analysis of Model Complexity}
We conduct experiments to analyze the computational complexity of our method from three aspects: window size for calculation of self-attention, overlapping cross-attention block (OCAB) and channel attention block (CAB). We also compare our method with the Transformer-based method SwinIR. The $\times$4 SR performance on Urban100 are reported and the number of Multiply-Add operations is counted at the input size of 64 $\times$ 64. Note that pre-training techniques (including $\times$2 pre-training) are \textbf{NOT} used for all the models in this section. The experimental setup is completely fair.

First, we use the standard Swin Transformer block as the backbone to explore the influence on different window sizes. As shown in Tab.~\ref{win_cmp}, enlarging window size can bring a large performance gain (+0.36dB) with a little increase in parameters and $\sim$$\%19$ increase in Multi-Adds.

\begin{table}[!t]
\center
\begin{center}
\vspace{6.1cm}
\caption{Model complexity comparison of window sizes.}
\label{win_cmp}
\setlength{\tabcolsep}{2.2mm}{
\begin{tabular}{c|ccc} 
\hline 
window size & $\#$Params. & $\#$Multi-Adds. & PSNR\\
\hline 
(8, 8) & 11.9M & 53.6G & 27.45dB \\
\hline 
(16, 16) & 12.1M & 63.8G & 27.81dB \\
\hline 
\end{tabular}
}
\end{center}
\end{table}

\begin{table}[!t]
\center
\begin{center}
\caption{Model complexity comparison of OCAB and CAB.}
\label{modules_cmp}
\setlength{\tabcolsep}{2.2mm}{
\begin{tabular}{c|ccc} 
\hline 
Method & $\#$Params. & $\#$Multi-Adds. & PSNR\\
\hline 
Baseline & 12.1M & 63.8G & 27.81dB \\
\hline 
w/ OCAB & 13.7M & 74.7G & 27.91dB \\
\hline 
w/ CAB & 19.2M & 92.8G & 27.91dB \\
\hline 
Ours & 20.8M & 103.7G & 27.97dB \\
\hline 
\end{tabular}
}
\end{center}
\end{table}

\begin{table}[!t]
\center
\begin{center}
\caption{Model complexity comparison of CAB sizes.}
\label{cab_cmp}
\setlength{\tabcolsep}{2.2mm}{
\begin{tabular}{c|ccc} 
\hline 
$\beta$ in CAB & $\#$Params. & $\#$Multi-Adds. & PSNR\\
\hline 
1 & 33.2M & 150.1G & 27.97dB \\
\hline 
2 & 22.7M & 107.1G & 27.92dB \\
\hline 
3 (default) & 19.2M & 92.8G & 27.91dB \\
\hline 
6 & 15.7M & 78.5G & 27.88dB \\
\hline 
w/o CAB & 12.1M & 63.8G & 27.81dB \\
\hline 
\end{tabular}
}
\end{center}
\end{table}

\begin{table}[!t]
\center
\begin{center}
\caption{Model complexity comparison of SwinIR and HAT.}
\label{method_cmp}
\setlength{\tabcolsep}{2.1mm}{
\begin{tabular}{c|ccc} 
\hline 
Method & $\#$Params. & $\#$Multi-Adds. & PSNR\\
\hline 
SwinIR & 11.9M & 53.6G & 27.45dB \\
\hline 
HAT-S (ours) & 9.6M & 54.9G & 27.80dB \\
\hline 
\hline 
SwinIR-L1 & 24.0M & 104.4G & 27.53dB \\
\hline 
SwinIR-L2 & 23.1M & 102.4G & 27.58dB \\
\hline 
HAT (ours) & 20.8M & 103.7G & 27.97dB \\
\hline 
\end{tabular}
}
\vspace{-0.3cm}
\end{center}
\end{table}

\begin{figure*}[!t]
\centering
\includegraphics[width=0.93\linewidth]{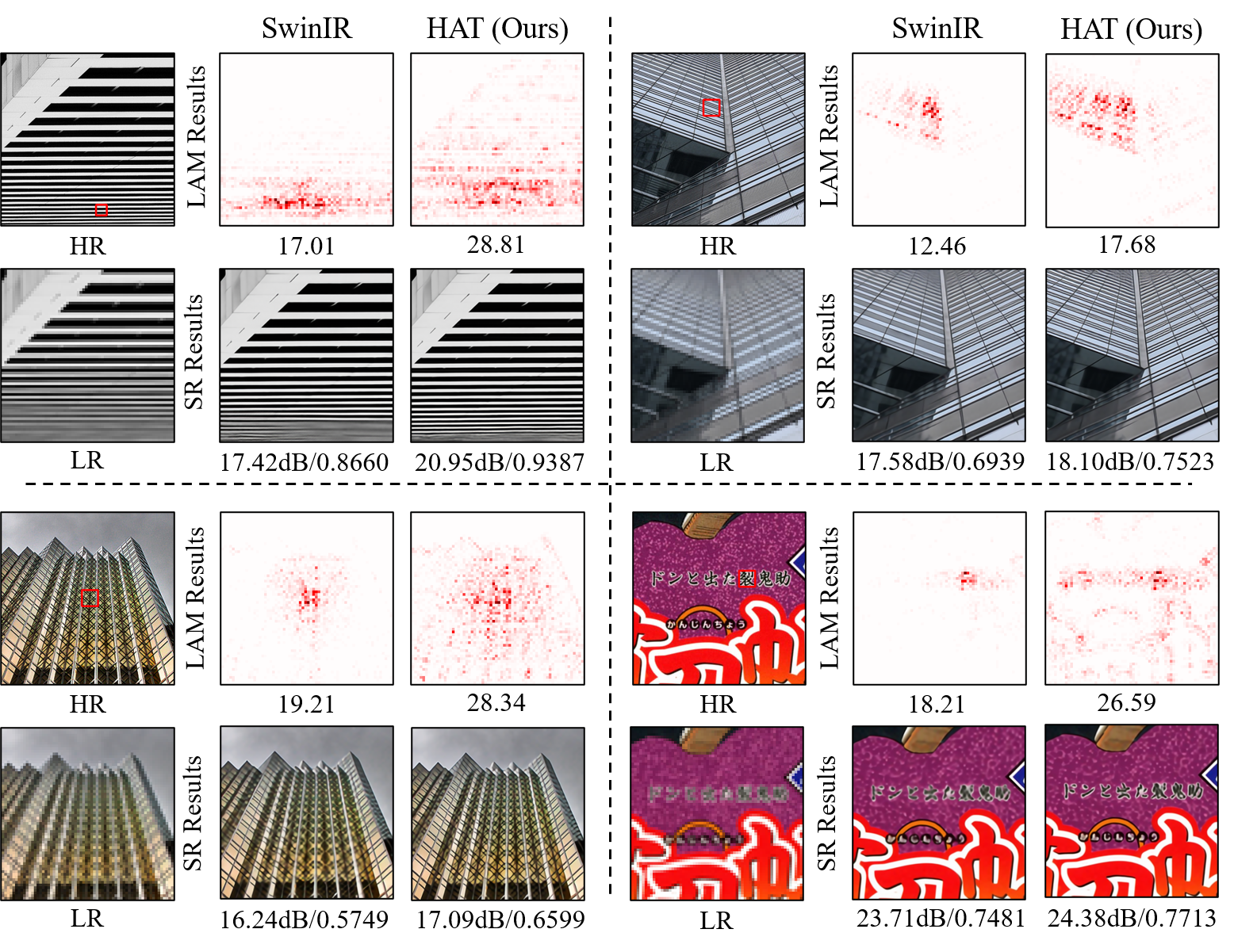}
\vspace{-0.3cm}
\caption{Comparison of LAM results between SwinIR and HAT.}
\label{more_lam}
\vspace{-0.4cm}
\end{figure*}

Then, we use window size 16 as the baseline to investigate the computational complexity of the proposed OCAB and CAB. As illustrated in Tab.~\ref{modules_cmp}, our OCAB obtains a performance gain with a limited increase of parameters and Multi-Adds. It demonstrates that the effectiveness and efficiency of the proposed OCAB. Besides, adding CAB to the baseline model also achieves better performance. 

Since CAB seems to be computationally expensive, we further explore the influence on CAB sizes by modulating the squeeze factor $\beta$ (mentioned in Sec.3.2.2 in the main paper). As shown in Tab.~\ref{cab_cmp}, adding a small CAB whose $\beta$ equals 6 can bring  considerable performance improvement. When we continuously reduce $\beta$, the performance increases but with larger model sizes. To balance the performance and computations, we set $\beta$ to 3 as the default setting.

Furthermore, we compare HAT and SwinIR with the similar numbers of parameters and Multi-Adds in two settings, as presented in Tab.~\ref{method_cmp}. 1) We compare HAT-S with the original version of SwinIR. With less parameters and comparable computations, HAT-S significantly outperforms SwinIR. 2) We enlarge SwinIR by increasing the width and depth of SwinIR to achieve similar computations as HAT, denoted as SwinIR-L1 and SwinIR-L2.
HAT achieves the best performance at the lowest computational cost. 

Overall, we find that enlarging the window size for the calculation of self-attention is a very cost-effective way to improve the Transformer model. Moreover, the proposed OCAB can bring an obvious performance gain with limited increase of computations. Although CAB is not as efficient as above two schemes, it can also bring stable and considerable performance improvement. Benefiting from the three designs, HAT can substantially outperforms the state-of-the-art method SwinIR with comparable computations. 

\section{More Visual Comparisons with LAM}
We provide more visual comparisons with LAM results to compare SwinIR and our HAT. The red points in LAM results represent the used pixels for reconstructing the patch marked with a red box in the HR image, and Diffusion Index (DI) is computed to reflect the range of involved pixels. The more pixels are utilized to recover the specific input patch, the wider the distribution of red points is in LAM and the higher DI is. As shown in Fig.~\ref{more_lam}, the LAM attribution of HAT expands to the almost full image, while that of SwinIR only gathers in a limited range. For the quantitative metric, HAT also obtains a much higher DI value than SwinIR. All these results demonstrate that our method activates more pixels to reconstruct the low-resolution input image. As a result, SR results generated by our method have higher PSNR/SSIM and better visual quality.

\end{document}